\newcommand{\bhar}{\ensuremath{\dot{M}_{\bullet}}\xspace}
\newcommand{\athenapk}{\texttt{AthenaPK}\xspace}

\newcommand{\train}{$t_{\mathrm{rain}}$\xspace}
\newcommand{\low}{\texttt{low-turb}\xspace}
\newcommand{\high}{\texttt{high-turb}\xspace}
\newcommand{\stormy}{\emph{stormy}\xspace}
\newcommand{\cloudy}{\emph{cloudy}\xspace}
\newcommand{\rainy}{\emph{rainy}\xspace}
\newcommand{\sunny}{\emph{sunny}\xspace}

\documentclass{aa}  
\usepackage{natbib}
\usepackage{hyperref}
\hypersetup{colorlinks=true,linkcolor=[rgb]{1.,0.2,0.2},citecolor=[rgb]{0.1,0.4,1.},filecolor=[rgb]{0.7,0.2,0.2},urlcolor=[rgb]{0.7,0.2,0.2}}
\usepackage{orcidlink}
\usepackage{graphicx}
\usepackage{txfonts}
\usepackage{booktabs}
\usepackage{amsmath}
\usepackage{amssymb}
\usepackage{verbatim}
\usepackage{longtable}
\usepackage{pdflscape}
\begin{document} 

\title{{\bfseries\scshape BlackHoleWeather –} Jet-regulated chaotic cold accretion across the meso scale: 
{\Large Morphology and thermodynamics}}

\titlerunning{Chaotic cold accretion in multiscale AGN feedback simulations}
\authorrunning{V.~Cammelli et al.}

\author{Vieri Cammelli \thanks{vieri.cammelli@unimore.it}
        \inst{1}\orcidlink{0000-0002-2070-9047},
        Massimo Gaspari
        \inst{1}\orcidlink{0000-0003-2754-9258},
        Olmo Piana
        \inst{1}\orcidlink{0000-0002-1558-5289},
        Filippo Barbani
        \inst{1}\orcidlink{0000-0002-1620-2577},
        Giovanni Stel
        \inst{1}\orcidlink{0009-0007-0585-9462},
        Davide M. Brustio
        \inst{1}\orcidlink{0009-0009-7700-1910},
        Valeria Olivares
        \inst{2,3}\orcidlink{0000-0001-6638-4324},
        Francesco Salvestrini
        \inst{4}\orcidlink{0000-0003-4751-7421},
        Ashkbiz Danehkar
        \inst{5}\orcidlink{0000-0003-4552-5997},
        Michael Reefe
        \inst{6}\orcidlink{0000-0003-4701-8497},
        Pasquale Temi
        \inst{7}\orcidlink{0000-0002-8341-342X},
        Filippo M. Maccagni
        \inst{8,9}\orcidlink{0000-0002-9930-1844},
        Francesco Tombesi
        \inst{10,11,12}\orcidlink{0000-0002-6562-8654}
        \and
        Martin Fournier
        \inst{13}\orcidlink{0009-0006-2593-1583}
        }

\institute{
Department of Physics, Informatics and Mathematics, University of Modena and Reggio Emilia, I-41125 Modena, Italy
\and
Department of Physics, Universidad de Santiago de Chile, Santiago, Chile
\and
CIRAS, Universidad de Santiago de Chile, Santiago, Chile
\and
INAF -- Osservatorio Astronomico di Trieste, Via G. Tiepolo 11, I-34143 Trieste, Italy
\and
Science and Technology Institute, Universities Space Research Association, Huntsville, AL 35805, USA
\and
Department of Physics and MIT Kavli Institute for Astrophysics and Space Research, Massachusetts Institute of Technology, Cambridge, MA 02139, USA
\and
NASA Ames Research Center, MS 245-6, Moffett Field, CA 94035-1000, USA
\and
INAF -- Osservatorio Astronomico di Cagliari, via della Scienza 5, 09047, Selargius (CA), Italy
\and
Wits Centre for Astrophysics, School of Physics, University of the Witwatersrand, 2000, Johannesburg, South Africa
\and
INAF -- Astronomical Observatory of Rome, 00078 Monte Porzio Catone (Rome), Italy
\and
Department of Physics, University of Rome ``Tor Vergata'', 00133 Rome, Italy
\and
INFN -- Rome ``Tor Vergata'' Section, 00133 Rome, Italy
\and
Universit\"{a}t Hamburg, Hamburger Sternwarte, Gojenbergsweg 112, 21029 Hamburg, Germany
}


\abstract
{How mechanical AGN feedback couples to multiphase condensation across scales remains a problem in galaxy groups and clusters. It is unclear how jets reshape the chaotic cold accretion (CCA) cycle and regulate black-hole fueling.}
{{\sc BlackHoleWeather} aims to build a unified description of the AGN baryon cycle across horizon, galactic, and group scales. Here we focus on how weather states shape the morphology and thermodynamics of jet-regulated CCA.}
{We perform two hydrodynamical simulations of a turbulent, radiatively cooling galaxy-group atmosphere with self-regulated AGN feedback. The runs are initialized in two turbulence regimes and evolved with a kinetic mass-loaded jet. Using 10 static refinement levels over a 100 kpc box, we reach sub-pc resolution, bridging the \textit{meso} scale down to below the Bondi radius.}
{The jet does not simply offset cooling via heating, but anisotropically reorganizes condensation through compression, shear, entrainment, and turbulent mixing. In the stronger-turbulence case ($\mathcal{M}\sim0.4$), condensation starts later ($t_{\rm rain}\simeq 16$ Myr) but becomes extended, filamentary, and mixed, with a broader hot-warm-cold bridge, a more porous cocoon, and burst-dominated fueling. At later times, this run evolves toward a cloud-dominated state with inefficient central accretion. In the weaker-turbulence case ($\mathcal{M}\sim0.15$), condensation starts earlier ($t_{\rm rain}\simeq 9$ Myr) and remains coherent and centrally confined, yielding a regular cocoon, a longer-lived inner cold reservoir with sustained fueling. In both runs, condensation is suppressed inside the jet channel and survives in the surrounding atmosphere and along the jet-ambient interface. Once condensation begins, SMBH fueling becomes super-Bondi.}
{These results extend CCA from a pure cooling$+$turbulence problem to a jet-regulated weather process. Ambient turbulence acts as a control parameter, producing an extended \stormy\ phase, a centrally retained \rainy\ cycle, and, in the high-turbulence case, a later \cloudy\ state with inefficient central fueling. The meso scale emerges as the layer linking halo thermodynamics to SMBH feeding within the broader {\sc BlackHoleWeather} framework.}

\keywords{black hole physics -- accretion, accretion disks -- hydrodynamics -- methods: numerical -- galaxies: active -- galaxies: groups: general}

\maketitle


\section{Introduction}\label{sec:intro}
Understanding the origin and evolution of multiphase outflows and relativistic jets from Supermassive Black Holes (SMBHs) remains a central unresolved problem in astrophysics. Within the theoretical framework of galaxy evolution, several arguments have been proposed to explain the origin of galactic-scale gas outflows through feedback processes powered by both stellar activity \citep{ChevalierClegg1985, VeilleuxCecil2005} and Active Galactic Nuclei (AGNs) \citep[e.g.][]{SilkRees1998, McNamaraNulsen2012, Fabian2012}. While stellar feedback represents an important channel \citep[e.g. ][]{BarbaniPascale2023,BarbaniPascale2025}, AGN feedback is often expected to dominate in massive galaxies found at the center of groups and clusters. In the AGN scenario, the energy released at the galactic center is primarily associated with gas accretion onto the central SMBH, located at the bottom of the gravitational potential well. This gives rise to a variety of processes collectively referred to as AGN feedback, which are believed to severely shape and affect the evolution of the interstellar, circumgalactic, and intracluster medium \citep[ISM, CGM, ICM; e.g.][]{Tumlinson2017, McNamaraNulsen2012}. Hence, AGN feedback may play a crucial role in regulating the formation and evolution of galaxies throughout the Universe. Among the most realistic and promising scenarios, AGNs launch fast winds linked to the central SMBH activity which in turn impact the surrounding medium \citep[e.g.][]{Menci2008, ZubovasKing2012, FaucherGiguere2012}. Despite considerable progress, a fully self-consistent theoretical framework describing the multiscale interplay between AGN activity and the surrounding galaxy or halo remains elusive. Depending on how the injected energy couples to the ambient multiphase medium, AGN feedback may either locally compress dense gas and promote star formation, or more globally suppress cooling and condensation through heating, evacuation, and turbulent stirring \citep[e.g.][]{SilkRees1998,GaspariBrighenti2011,Gaspari2012,Gaibler2012,Zubovas2013,Harrison2017,Morganti2017}.
The net outcome likely depends on how feedback couples to a turbulent, radiatively cooling atmosphere: AGN activity can both suppress global cooling and locally promote nonlinear condensation through uplift, compression, and mixing, thereby establishing a self-regulated multiphase cycle rather than a purely one-way heating process.

In the {\sc BlackHoleWeather} framework \citep{Gaspari2020}, AGN feeding and feedback are treated as a multiscale baryon cycle analogous to atmospheric weather: turbulent hot halos develop localized nonlinear cooling, which seeds a top-down condensation cascade from the X-ray phase to warm filaments and cold clouds; these multiphase structures then rain toward the SMBH, where chaotic collisions promote angular-momentum cancellation and strongly variable fueling. The resulting accretion powers mechanical feedback, which in turn stirs, uplifts, compresses, and reheats the surrounding atmosphere, thereby reshaping the conditions for the next condensation episode. The key challenge is therefore not to understand cooling or heating in isolation, but to capture the full self-regulated weather cycle across the \textit{macro}- ($r\gtrsim1$\,kpc), \textit{meso}- ($0.1\lesssim r\lesssim 1$\,kpc), and micro- ($<0.1$ kpc) scales that connect the virialized halo to the SMBH vicinity with particular emphasis on the transition between large scale rain/ejecta to inner inflow/outflow properties through the meso-scale.

In galactic and AGN environments, the study of the processes driving the gas across different scales and phases is pivotal as simulations currently lack of exploration in the intermediate meso-scale regime. At \textit{macro} scales, the hot diffuse gas traces the energy deposited by AGN and stellar feedback, manifested through shocks, cavities, mixing, and turbulent cascades driven by jets and winds. Localized thermal instabilities occurring on \textit{meso} scales trigger initial condensation of hot to warm gas, observed in the optical/UV phase while rapidly cooling. Eventually, cold gas typically assembles in dense clumps and filaments (\textit{micro} scale, $r\lesssim0.1$\,kpc), where the gas has undergone significant radiative cooling, providing fuel for subsequent SMBH accretion and/or star formation. The largest challenge to overcome in understanding the accretion and outflow-driven feedback loop in the dominant galaxy of groups and clusters of galaxies is the enormous dynamic range in spatial and mass scales which must be considered. In fact, while galaxy groups and clusters extend further out to the Mpc scale, the physics of the actual SMBH gas accretion and subsequent jet injection take place well below parsec scales, eventually reaching the BH horizon/Schwarzschild scale (about $\sim 10^{-5}$ pc for a SMBH of mass $10^{8}\ M_\odot$). Within $\sim 100$ Schwarzschild radii ($r_{\rm S}$) from the black hole, only a small fraction of the infalling, or `raining' gas is actually accreted by the black hole \citep[e.g. ][]{Gaspari2017}. The gravitational binding energy of this gas is partially converted into mechanical energy, manifesting as either ultrafast outflows \citep[with velocities on the order of $0.1\ c$, see, e.g. ][]{Tombesi2010, TombesiCappi2011, TombesiCappi2013} or highly collimated relativistic jets \citep[see ][for a review]{BlandfordMeier2019}, particularly in cases involving rapidly spinning SMBHs. These mechanisms are thought to initiate the AGN feedback phase \citep{Fabian2012}. Additionally, radiation emitted from the accretion disk regions ignites another feedback channel, especially significant at high Eddington ratios $\gtrsim0.1$ \citep[e.g. ][]{Ciotti2007}.

In terms of morphology studies, ultraviolet imaging of brightest cluster galaxies reveals star formation in clumps, knots, and filaments, often extending to radii of $\sim 10-30$ kpc, consistent with multiphase condensation and with star formation triggered or enhanced by AGN-driven uplift and jets \citep[e.g. ][]{Donahue2015,Fogarty2015,Reefe2025}. Likewise, millimeter and sub-millimeter observations show cold molecular filaments whose morphology often follows radio jets \citep[e.g.][]{Russell2019,OlivaresSalome2019,Castignani2025}. These structures are generally interpreted as cold gas condensing from the hot atmosphere, redistributed by AGN feedback, and in some cases forming stars in situ. Similar alignments are also found at high redshift, where ALMA detects extended molecular reservoirs in the CGM co-spatial with radio jets, supporting a close connection between AGN activity and cold-gas condensation \citep[e.g.][]{LiEmonts2021,WalterBanados2025}.

On top of this, in the last few decades X-ray observations have unmistakably confirmed that AGNs recursively perturb the surrounding gaseous environments via shocks, bubbles, and outflows, with energies as large as $10^{46}$ erg s$^{-1}$ detected with Chandra and XMM-Newton \citep{Fabian2012, GittiBrighenti2012, McNamaraNulsen2012}. Most recently, AGN astronomy has witnessed a surge in the detection of multiphase AGN outflows, marking a ``golden'' age of observational discovery. High-resolution spectroscopy and multi-wavelength imaging, primarily via ALMA, MUSE, JWST, XMM-Newton and XRISM, have revealed increasingly detailed properties of these phenomena \citep{Alatalo2013, Cicone2014, Fiore2017, Mainieri2021, Maiolino2017, Morganti2017, Tombesi2010, Veilleux2020, XrismCollaborationAudard2025}. Observed outflows exhibit a continuum of phases, with mass loss rates that span $\dot{M}_{\rm out} \sim 0.1 - 100\ M_\odot\ \text{yr}^{-1}$ and velocities ranging from $v_{\rm out} \sim 10^2$ to $10^4$ km\ s$^{-1}$, extending over the ionized to the molecular phase \citep{KingPounds2015, TombesiCappi2013, TombesiMelendez2015}. Despite this progress, a fully self-consistent theoretical framework describing how mass and energy are transported from the event horizon to kpc scales remains elusive.

Within the current theoretical {\sc BlackHoleWeather} framework, \emph{chaotic cold accretion} (CCA) has emerged as a particularly promising scenario for connecting halo thermodynamics, multiphase structure, SMBH fueling, and their observable kinematic and temporal signatures. In this picture, hot turbulent atmospheres, despite remaining in approximate global thermal equilibrium, develop localized thermal instabilities that drive the nonlinear condensation of cold clouds and filaments. These structures repeatedly collide, efficiently cancel angular momentum, and eventually `rain' toward the SMBH, giving rise to stochastic accretion and intermittent feedback \citep{Gaspari2013,Gaspari2015,Gaspari2017}. CCA therefore predicts not only the widespread presence of multiphase gas, but also distinct variability and kinematic diagnostics, since correlated inflow/outflow cycles can increase the velocity dispersion of the multiphase medium, broaden line profiles, and strongly modulate the BH accretion rate and jet power.
More broadly, CCA provides a physically motivated framework linking the thermodynamic state of hot halos to the emergence of multiphase gas and the feeding of the central AGN. This interpretation is supported by a growing body of observations across the radio, millimeter, optical, ultraviolet, and X-ray bands \citep[e.g. ][]{Voit2015,Mcdonald2018,Tremblay2018,Maccagni2021,Temi2022,OlivaresSalome2022,Olivares2025,Reefe2025,Romero2025,Omoruyi2026}.

At the same time, preliminary studies employing general relativistic magnetohydrodynamic (GRMHD) simulations \citep{Sadowski2017, Tchekhovskoy2012} have begun to constrain both the mechanical and radiative efficiencies near the event horizon, although within limited computational domains and simplified plasma physics treatments. Semi-analytical models predict that ultrafast nuclear outflows will gradually interact with the surrounding interstellar medium through shock fronts and hydrodynamic instabilities \citep[e.g. ][]{Fiore2017, MenciFiore2019}. This interaction leads to the entrainment of ambient material, which slows the outflow and causes cooler and broader components such as neutral and molecular winds \citep{FaucherGiguere2012, Wagner2013, Zubovas2014}.  

This work is embedded within the {\sc BlackHoleWeather} project (PI: Gaspari), whose goal is to build a unified, self-consistent, multi-physics framework for SMBH feeding and feedback starting from first principles \citep{Gaspari2020}. Within this perspective, the AGN phenomenon is viewed as a genuinely multiscale baryon cycle, linking halo-scale cooling and condensation to circumnuclear accretion and the subsequent release of mechanical feedback. The present study represents a key step in this broader program and, in particular, contributes to WP3, the work package dedicated to meso- to micro-scale \emph{feedback}. \cite{Barbani2026a, Barbani2026b}, hereafter B26a,b, present the numerical setup and baseline implementations in a purely turbulent atmosphere without AGN feedback, demonstrating how radiative cooling alone can produce different BH `weather' regimes across scales. Several complementary {\sc BlackHoleWeather} studies are now extending this framework toward additional key physical ingredients. In particular, \cite{Piana2026a, Piana2026b}, hereafter P26a,b, will examine the role of black hole spin, accretion, angular momentum transport, and relativistic effects, while other ongoing efforts investigate the influence of dust and chemistry on multiphase accretion flows.

The presented study investigates the impact of AGN feedback on the surrounding gas in galaxy-group environments. To this end, we perform consistent hydrodynamical simulations spanning from sub-pc scales in the immediate SMBH vicinity up to group scales of hundreds of kpc, following the evolution of multiphase gas under the combined effects of gravity, radiative cooling, turbulence, jet launching, and mechanical energy deposition. Within \athenapk, we employ static mesh refinement (SMR) to zoom down to sub-parsec scales, enabling a detailed treatment of SMBH accretion and jet-driven feedback in both the interstellar and intragroup media, while building on the controlled stratified-halo setups that established the CCA framework \citep{Gaspari2012,Gaspari2013,Gaspari2017}. Akin to B26a,b, the analysis is divided into two companion papers. In the present paper (C26a), we focus on the morphological and thermodynamic properties of the multiphase gas in the context of CCA, and on how AGN feedback reshapes the medium through its impact on turbulence, cooling, and the spatial and temporal distribution of the gas. The companion paper  \cite{Cammelli2026b}, hereafter C26b, instead examines the CCA diagnostics \citep{Gaspari2018}, focusing on the kinematics and time variability extracted from the same simulation suite.

The structure of the paper is as follows. \S\ref{sec:phys} introduces the physical basis of gas inflow, outflow, and subgrid feedback energetics. The numerical setup and reference runs are described in \S\ref{sec:setup}. In \S\ref{sec:results}, we present the main simulation results, focusing on the morphology, thermodynamics, phase structure, and SMBH fueling history of the jet-regulated CCA flow. In \S\ref{sec:disc}, we discuss the physical implications of these results, define the operational weather stages (\sunny, \stormy, \rainy, and \cloudy), and compare the emerging jet-regulated sequence with previous simulation work and companion BHW studies. Finally, our conclusions are summarized in \S\ref{sec:concl}.

\section{Numerical methods}\label{sec:setup}
Our simulations are performed via the \athenapk\footnote{https://github.com/parthenon-hpc-lab/athenapk} code for astrophysical magneto-hydrodynamics \citep{Stone2020, Grete2023}. 
\athenapk advances the field by solving the full set of MHD equations at each timestep while being explicitly designed for the era of parallel GPU computing. Unlike the majority of existing astrophysical simulation codes, which were developed prior to the widespread adoption of GPUs, \athenapk has been rigorously tested and shown to scale linearly on several NVIDIA architectures. This capability, combined with the GPU resources available on the Leonardo-Booster partition, is essential for achieving the unprecedented spatial resolution and physical fidelity required by the BHW project. \athenapk is built on the well-validated \texttt{Athena++} framework and extended through the \texttt{Parthenon} \citep{Grete2023} and \texttt{Kokkos} \citep{Edwards2014} libraries, ensuring both portability and scalability across current and next-generation HPC systems. In this design, \texttt{Parthenon} manages adaptive mesh refinement (AMR), while \texttt{Kokkos} provides highly efficient on-node data parallelism. Inter-node communication is optimized via \texttt{Parthenon}’s asynchronous one-sided MPI calls, with data transferred directly from GPU memory buffers, avoiding costly bottlenecks.

Progressing in this field requires exceptional computational resources as the physical problem embeds a huge range of dynamical scales, and heavy parallel workflows have become a necessary requirement in the recent past. In particular, GPUs (Graphical Processing Units) implementation is paving the way towards the era of High Performance Computing (HPC), as it has been shown that their usage boosts the computational capabilities of already existing astrophysical codes by at least a factor of $\sim10-100$. In terms of portable GPU-accelerated hydrodynamical frameworks, several approaches have emerged in recent years. AMReX \citep{Zhang2019} provides a highly flexible, though complex, mesh infrastructure and implements on-node parallelization natively, rather than through an external library. UINTAH \citep{Holmen2017} similarly adopts \texttt{Kokkos} for performance portability, but unlike \texttt{Parthenon} , executes tasks at a finer granularity, which improves flexibility and scalability at the expense of requiring many more kernel launches per device. CHARM++ has also incorporated GPU support \citep{Choi2022} and supports QUINOA \citep{Bakosi2017}, while GAMER-2 \citep{Schive2018} represents a GPU-accelerated astrophysical AMR framework, although its reliance on CUDA alone limits portability, and its host–device memory design introduces significant transfer overheads. 


For the complete description of the simulation setup, including the implementation of the SMBH sink region at the centre of the box, radiative cooling, turbulence driving and initial conditions, we refer to B26a. In the following we focus on highlighting the key aspects of the simulations, discussing in detail the major setup differences compared with B26a and describing the novel physics added via basic principles. As mentioned in the previous Section, for the sake of this study we ignore the presence of magnetic fields, and we defer the study of how they would affect the evolution of such systems in a future paper of BHW series. 

\subsection{Feedback prescriptions and subgrid energetics}\label{sec:phys}

Because the sink region does not resolve the micro accretion disk or the Innermost Stable Circular Orbit (ISCO, where $r_{\rm ISCO}\sim3-9\times r_{\rm S}$), we adopt a subgrid mapping between the sink accretion rate and the SMBH feeding rate. In this paper the efficiencies are kept fixed and spin-independent, and the resulting SMBH accretion rate is used to normalize the feedback power rather than to model disk microphysics in detail.

Since the simulation outputs the local accretion rate onto the sink particle as finest quantity, at the occurrence of every accretion episode onto the sink $\dot{M}_{\rm sink}$, a corresponding accretion rate onto the SMBH \bhar is associated as follows:
\begin{equation}\label{eq:bhar}
    \dot{M}_{\bullet} = \big(1 - \epsilon_{\rm disk}\big) \dot{M}_{\rm sink},
\end{equation}
where $\epsilon_{\rm disk}$ is the efficiency of the accretion disk in dissipating the gravitational energy of a gas particle from the sink radius down to the ISCO radius. Assuming that the dissipated energy is converted mainly into radiation, one can think of $\epsilon_{\rm disk}$ as the radiative efficiency of the accretion disk such that
\begin{equation}\label{eq:eff_isco}
    \epsilon_{\rm ISCO} = 1 - \epsilon_{\rm disk} \equiv 1 - \epsilon_{\rm rad,disk}.
\end{equation}
In the above expression $\epsilon_{\rm ISCO}$ represents the dimensionless mass-energy budget of a test particle calculated at the ISCO radius relative to its total energy at the sink radius. We note here that a priori the actual efficiencies introduced above will also depend on the current BH spin and the specific accretion state of the disk. Under the assumption of a radiatively efficient disc, the dissipated gravitational energy is mainly emitted as radiation, so that $\epsilon_{\rm disk}$ can be interpreted as the disc radiative efficiency, $\epsilon_{\rm rad,disk}$. In the standard thin-disc picture, this efficiency is set primarily by the binding energy released down to the ISCO and is therefore spin-dependent, ranging from a few percent for a non-rotating BH up to $\sim 0.3$--$0.4$ for a rapidly spinning prograde Kerr BH \citep[for a review, see ][]{AbramowiczFragile2013}. We defer the study of this case to companion works (see P26a,b), while in this work we will focus on the impact of the BH feedback keeping constant efficiencies.  

In the general case, the relative fractions of energy injected at the micro scale, being it kinetic, thermal or magnetic, define the physical mechanism driving the AGN feedback. We note that in this work there is no formal distinction between jets and winds, and the implemented jet is meant to be a pure mechanical outflow (see \S\ref{subsec:jet}).  Assuming the usual mass-energy relativistic conversion and that the magnetic energy only contributes to the local density and pressure, one can neglect the magnetic term and express the total feedback power $P_{\rm tot}$ as
\begin{equation}\label{eq:tot_power}
    P_{\rm tot} = P_{\rm kin} + P_{\rm th} = \epsilon_{\rm tot} \dot{M}_{\bullet} c^{2} \equiv \big(\epsilon_{\rm kin} + \epsilon_{\rm th}\big) \dot{M}_{\bullet} c^{2}, 
\end{equation}
where we introduce the total efficiency $\epsilon_{\rm tot}$ as the sum of the jet $\epsilon_{\rm kin}$ and thermal $\epsilon_{\rm th}$ efficiencies. Please note that in the majority of cases one can approximate that most of the thermal energy not injected via jet is radiated away, i.e. $\epsilon_{\rm th}\simeq\epsilon_{\rm rad}$ (this radiative efficiency is different from the one introduced in Eq.~(\ref{eq:eff_isco})). 

The outflow mass loading is prescribed as
\begin{equation}\label{eq:mdot_out}
\dot M_{\rm out}=\eta_{\rm jet} \dot M_\bullet,
\end{equation}
where $\eta_{\rm jet}$ is the dimensionless mass-loading factor of the subgrid jet. 
Motivated by GRMHD results, in which a large fraction of the accreting gas reaching the nuclear region is often re-ejected back by the SMBH \citep[e.g.][]{GaspariSadowski2017}, we adopt the simple reference closure $\eta_{\rm jet}=1-\epsilon_{\rm tot}$. This ties the injected jet mass directly to the SMBH feeding rate, while the feedback power is independently normalized by $P_{\rm tot}=\epsilon_{\rm tot}\dot M_\bullet c^2$ \citep[see also ][]{GreteO'Shea2025}.

In this work, we adopt a fixed total feedback efficiency of $\epsilon_{\rm tot}=0.028$, motivated by the GRMHD simulations of \citet{Sadowski2017} and assuming on average a low spin BH ($a\simeq0.2$), which indicates that mechanically dominated SMBH outflows can channel a few per cent of the accreted rest-mass energy into feedback. We use this value here as a representative subgrid normalization for a kinetic jet, while keeping it fixed in order to isolate the impact of the ambient turbulence level and the jet-atmosphere coupling on the multiscale evolution. A broader exploration of the dependence on spin and efficiency is deferred to companion works (i.e., P26a,b). If the velocity of the gas is mainly driven by the jet kinetic and thermal energy contribution, one can work out the initial velocity of the jet gas at the injection site as
\begin{equation}\label{eq:approx_tot_power}
    P_{\rm tot} \simeq P_{\rm kin} + P_{\rm th, jet} = \epsilon_{\rm tot} \dot{M}_{\bullet} c^{2},
\end{equation}
where we assume that most of the thermal energy is carried by the jet.
We can now apply the definition of kinetic and thermal energy such that
\begin{equation}\label{eq:kin_power}
    P_{\rm kin} = \frac{1}{2} \dot{M}_{\rm jet} v_{\rm jet}^2,
\end{equation}
where $v_{\rm jet}$ is the velocity of the ejected material at the injection site, and
\begin{equation}\label{eq:th_power}
    P_{\rm th,jet} = e_{\rm jet} \dot{M}_{\rm jet},
\end{equation}
where the internal energy per unit mass of the jet gas is denoted by $e_{\rm jet}=k_{\rm B}T_{\rm jet}/[(\gamma-1)\mu m_{\rm p}]$, with \(T_{\rm jet}=5\times10^8\,{\rm K}\). Plugging Eqs.~(\ref{eq:kin_power}) and (\ref{eq:th_power}) in Eq.~(\ref{eq:approx_tot_power}) gives
\begin{equation}
    \epsilon_{\rm tot} \dot{M}_{\bullet} c^{2} = \Big(\frac{1}{2}v_{\rm jet}^2 + e\Big) \dot{M}_{\rm jet}. 
\end{equation}
Using Eq.~(\ref{eq:mdot_out}) and solving for $v_{\rm jet}$ leads us to
\begin{equation}\label{eq:v_out}
    v_{\rm jet} = \sqrt{\frac{2}{(1-\epsilon_{\rm tot})}\Big[\epsilon_{\rm tot}c^2 - (1-\epsilon_{\rm tot}) e_{\rm jet}\Big]}.
\end{equation}
With the adopted fiducial feedback efficiency above, this yields \(v_{\rm jet}\simeq0.25\,c\) in the simulations presented here.

\subsection{Grid skeleton}
For the scope of this paper, we perform \athenapk runs simulating a $100$ kpc per side volume using a Static Mesh Refinement (SMR) grid at 10 refinement levels (see \S\ref{subsec:sim_suite} for the full details of the reference runs). The root grid has $128$ cells per side, where every 32 cells define a mesh block resulting in $4^3$ total mesh blocks. This sets the coarsest resolution element to be $\Delta x_{\rm root}\simeq 0.78$ kpc. Any extra refinement level is added by refining the innermost layer of mesh blocks around the origin of the box, namely halving each cell side, as shown in Figure~\ref{fig:grid}. Hence, considering equal volumes, the refinement region contains a number of cells 8 times larger than the coarser volume element. Bearing this in mind, the finest resolution at the $n$-th refinement level corresponds to $\Delta x_{n} = \Delta x_{\rm root}/2^n$, being equivalent to $\simeq 0.78$ pc at $10$ refinement levels. 

We adopt static rather than adaptive refinement because the central region is continuously relevant throughout the simulation: it contains the sink, the jet injection zone, and the steepest thermodynamic gradients associated with cooling, condensation, and feedback coupling. In this context, a fixed refinement hierarchy offers two practical advantages. First, it ensures a reproducible and stable numerical environment for the sink-jet interaction, avoiding spurious fluctuations that could arise from repeated refinement and de-refinement near the SMBH. Second, it provides a more regular workload and memory pattern for GPU execution, which is advantageous for performance and for clean run-to-run comparisons. Since our main goal is to isolate the physical impact of ambient turbulence on multiscale jet-regulated condensation, we prefer a controlled SMR setup over a dynamically evolving mesh geometry.

\subsection{Hydrodynamics}\label{subsec:hydro}
The gas hydrodynamics in each cell are evolved in a conservative Eulerian fashion by integrating the usual system of equations:
\begin{equation}
\dfrac{\partial \rho}{\partial t} + \nabla \cdot (\rho \boldsymbol{v})= \mathcal{S}^{\text{jet}}_{\rho},
\end{equation}

\begin{equation}
\dfrac{\partial (\rho \boldsymbol{v})}{\partial t} + \nabla \cdot (\rho \boldsymbol{v} \otimes \boldsymbol{v} + P\mathbb{I})= -\rho \nabla \Phi + \rho \boldsymbol{f}_{\text{turb}} + \boldsymbol{\mathcal{S}}^{\text{jet}}_{\rho\boldsymbol{v}},
\end{equation}

\begin{equation}
\dfrac{\partial E}{\partial t} + \nabla \cdot ( E + P ) \mathbf{v}= -\rho \mathbf{v} \cdot \nabla \Phi - \mathcal{L} + \mathcal{S}_{\text{turb}} + \mathcal{S}^{\text{jet}}_{E},
\end{equation}
where we depict $\rho$ as the mass density, $\mathbf{v}$ as the velocity, $P$ as the gas pressure and $E$ as the total energy of the cell defined as $E=\rho e+\rho v^2/2$. The source term $\Phi$ represents the total gravitational potential, while cooling losses are described via $\mathcal{L}=n^2 \Lambda(T)$, which in turn is a function of the cooling function $\Lambda$ (see B26a for details) and of the number density of the gas $n$. Turbulence injection is introduced as usual by the energy deposited per unit time $\mathcal{S}_{\text{turb}}= \rho \boldsymbol{v} \cdot \boldsymbol{f}_{\text{turb}}$, with $\boldsymbol{f}_{\text{turb}}$ being the typical acceleration due to turbulent motion injected at a given scale (see B26a for turbulence driving details). The continuity, momentum and energy jet terms, $\mathcal{S}^{\text{jet}}_{\rho}$, $\boldsymbol{\mathcal{S}}^{\text{jet}}_{\rho\boldsymbol{v}}$ and $\mathcal{S}^{\text{jet}}_{E}$, are added according to a \textit{mass-loaded} jet along a given axis and are described in detail in the next section. The gas is closed with an ideal monatomic equation of state, with adiabatic index $\gamma = 5/3$.

The numerical approach is a second-order accurate, finite-volume Godunov scheme \citep{Godunov1959}. We achieve spatial accuracy using the Piecewise Linear Method (PLM) for reconstruction and temporal accuracy via a second-order Runge--Kutta scheme \citep{Butcher2008} for time integration. Fluxes at cell interfaces are determined by solving the Riemann problem with the HLLC approximate solver \citep{Toro1994}, chosen for its robust handling of contact and shear discontinuities. A first-order flux correction \citep{Bruggen2023} is applied locally near strong discontinuities to enhance numerical stability and accuracy in complex flows. For gas cells that exhibit unphysical states (e.g. negative densities $\rho < 0$ or temperatures $T < 0$), a robust fallback procedure is automatically triggered, which does not significantly impact the global energy conservation budget. This involves recalculating the fluxes using a simpler, inherently stable combination: forward Euler time integration, first-order (piecewise constant) reconstruction, and the Local Lax--Friedrichs Riemann solver. This ensures stable, physically sound solutions without imposing artificial floors on the thermodynamic variables. We note here that the only conditions we enforce concern imposing positive values of density and pressure, with a temperature floor down to 1 K. 

\subsection{Micro-scale jet}\label{subsec:jet}

\begin{figure}
    \centering
    \includegraphics[width=1\linewidth]{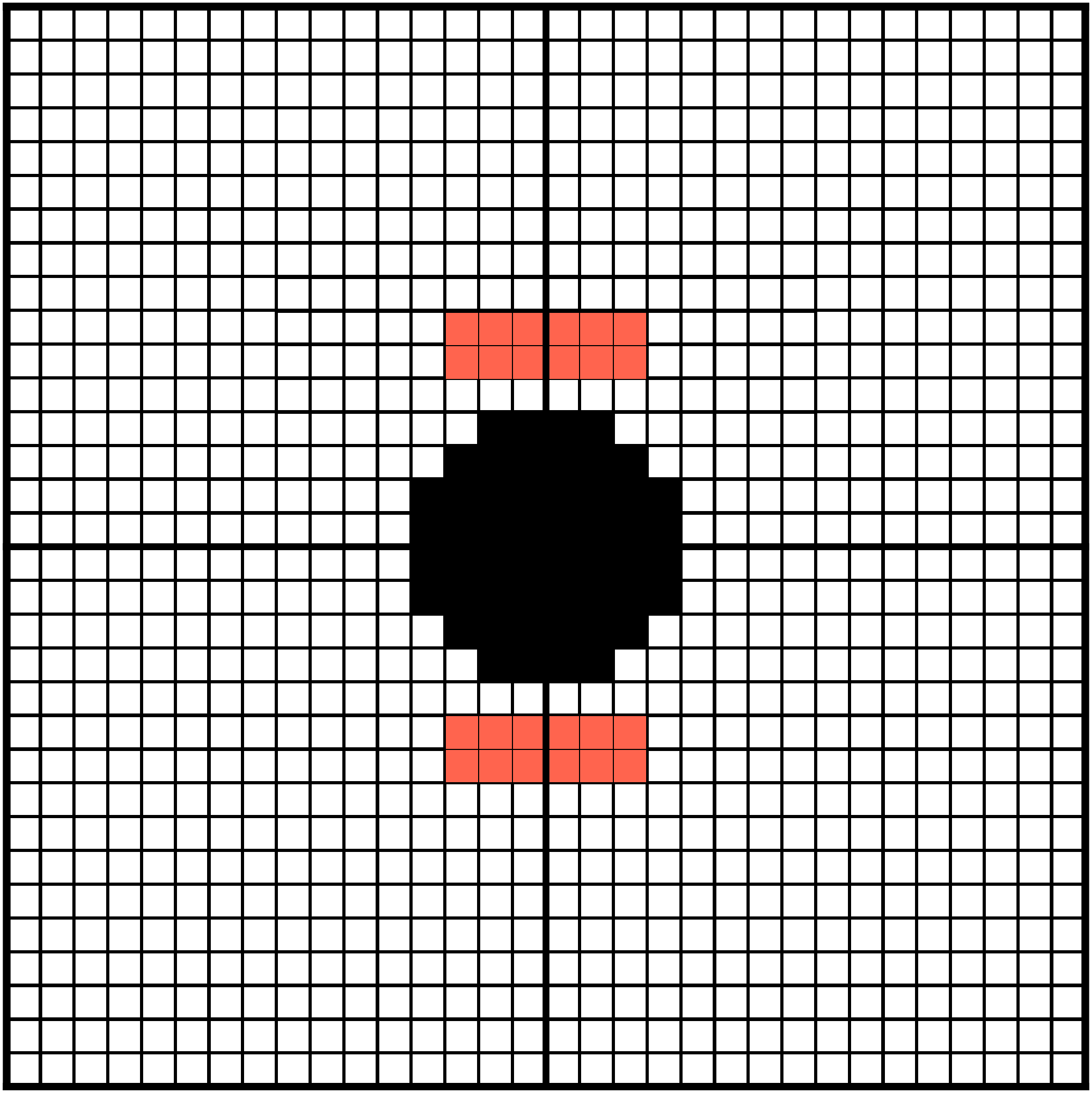}
    \caption{SMR grid visualization showing the sink particle (black cells) and the jet injection zone (orange cells) at the innermost refinement levels. At each iteration, the gas density, the temperature, and the velocity are reset to $\rho_{\rm sink} = 10^{-30}$ g\,cm$^{-3}$, $T_{\rm sink}=1$, and $v_{\rm sink}=0$ km\,s$^{-1}$, respectively. The radius of the sink particle $r_{\rm sink}$ is expressed in cell length units and has been set to 4 cells. Figure adapted from B26a. }
    \label{fig:grid}
\end{figure}

Within the main environment of \athenapk, we develop a dedicated feedback module which incorporates the physics of the jet at \textit{micro} scales (sub-parsec) \citep{Fournier2024,GreteO'Shea2025,PrasadGrete2026}. As a baseline, we start by assuming that the SMBH and its vicinity can be modeled by a spherical sink region centered at the origin, where the SMBH is located (see \citealt{Gaspari2013} and B26a for details) at the highest refinement level, as depicted by the black cells in Figure~\ref{fig:grid}. At each iteration, the gas density, the temperature, and the velocity are reset to $\rho_{\rm sink} = 10^{-30}$ g\,cm$^{-3}$, $T_{\rm sink}=1\ K$, and $v_{\rm sink}=0\ $km\,s$^{-1}$, respectively. The radius of the sink particle $r_{\rm sink}=3.12$ pc is expressed in cell length units ($\Delta x_{\rm min}\simeq0.78$ pc) and has been set to 4 cells for all our production runs. This value has been tested over several configurations to ensure a stable setup for both the physical and computational evolution.

The micro-feedback routine injects mass, momentum, and energy into a prescribed jet region in a manner consistent with the global accretion power budget. The jet-driven feedback consists of a pair of oppositely directed, collimated cylinders centered on the accreting object. Each jet is defined geometrically by a cylindrical disk of radius $r_{\rm jet}\simeq2.34$ pc and thickness $d_{\rm jet}\simeq1.56$ pc along the $\pm z$--axis, such that the injection volume approximates the base of a bipolar outflow attached to the sink, as shown in Figure~\ref{fig:grid}. 

At each timestep, the mass injected into the bipolar outflow is set by the subgrid mass-loading relation (Eq.~\ref{eq:mdot_out}), while the associated feedback power is normalized by
$P_{\rm tot}=\epsilon_{\rm tot}\dot M_\bullet c^2$. The injected momentum and energy are then added along the jet axis consistently with the velocity derived in Eq.~\ref{eq:v_out}. Within the injection region, the conserved hydrodynamical variables are updated according to
\begin{align}
    & \mathcal{S}^{\text{jet}}_{\rho} = \dot{\rho}_{\rm jet} \, (W_+ +W_-), \\
    & \boldsymbol{\mathcal{S}}^{\text{jet}}_{\rho\boldsymbol{v}} = \dot{\rho}_{\rm jet} \, v_{\rm z} \boldsymbol{\hat{z}}\, W_{+} - \dot{\rho}_{\rm jet} \, v_{\rm z} \boldsymbol{\hat{z}}\, W_{-}, \\
    & \mathcal{S}^{\text{jet}}_{E} = \dot{\rho}_{\rm jet} \, \Big( e_{\rm jet} +  \tfrac{v_{z}^{2}}{2}\Big) \, (W_+ +W_-). 
\end{align}
where $\dot{\rho}_{\rm jet}$ is the mass injection rate per unit volume, $v_{z}=v_{\rm jet}$ is the injected velocity (Eq.~(\ref{eq:v_out})), and $e_{\rm jet}$ the internal energy per unit mass corresponding to the chosen jet temperature (see also Eq.~(\ref{eq:th_power})). 
We note that the normalization of the jet mass source, $\dot{\rho}_{\rm jet}$, is fixed by the prescribed outflow rate in Eq.~(\ref{eq:mdot_out}), after accounting for the total injection volume and the normalization of the spatial kernels.
For the sake of formalism, we introduce the functions $W_+$ and $W_-$ as normalized spatial kernels that distribute the source terms within the injection volume. We note that for the scope of this work, the jet is always assumed to be co-aligned with the $\pm z$--axis, with no opening angle (hence $|W_{\pm}|\simeq1$). 
In this paper, the jet is injected along the $\pm z$ axis with no imposed opening angle, in order to minimize subgrid geometric assumptions and isolate the role of ambient turbulence in shaping the subsequent flow. The impact of a time-varying jet reorientation is explored separately in P26a,b.

The input of associated energy is determined by the available feedback power $P_{\rm tot}$ by splitting the total budget between the kinetic and thermal channels according to the input fractions via Eq.~(\ref{eq:approx_tot_power}). In the most general case, the kinetic channel deposits momentum at a fixed jet velocity, while the corresponding energy increment is applied to the total energy; this ensures that the kinetic energy implied by the momentum addition is explicitly represented. In the thermal channel, an isotropic internal energy contribution is injected either within the same collimated region or in a surrounding spherical bubble, depending on parameter choice. The scheme thus maintains global energy conservation while allowing the relative balance of kinetic and thermal feedback to be controlled via input parameters, and it naturally enforces the consistency between injected momentum and energy that is required to avoid unphysical negative pressures in the simulation cells.

\begin{figure*}
    \centering
    \includegraphics[width=0.8\linewidth]{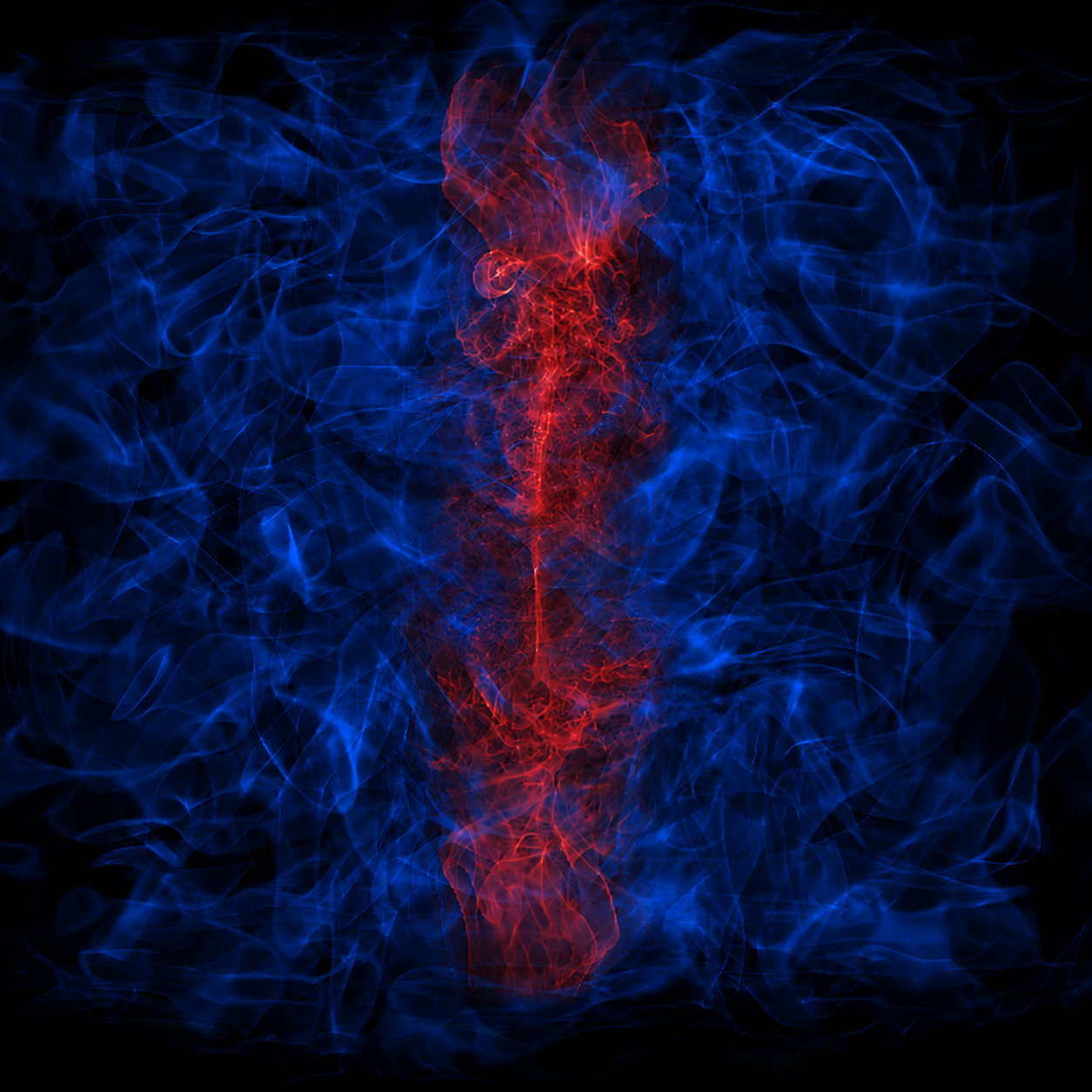}
    \caption{Volume rendering of the jet injected at the sink level observed from a location outside the simulation box. Blue color represents the turbulent gas, mainly introduced by the ad-hoc turbulence driving. Reddish gas is instead jet entrained material, while the white depicts the jet core which is expelled at the injection region sitting on the sink.}
    \label{fig:jet_rendering}
\end{figure*}

A visual rendering of the jet material injected in the simulated box is shown in Figure~\ref{fig:jet_rendering}. Here we show a 3D volume rendering of the simulation box, where the blue color traces the turbulent gas driven as described in the previous Sections (see B26a for details). The entrained material along the jet is instead seen in red colors, while the jet core is displayed as a high-intensity white region.

\subsection{Initial conditions \& Simulation runs}\label{subsec:sim_suite}
Our simulation design builds on the controlled stratified-halo setups that established the CCA framework \citep{Gaspari2012,Gaspari2013,Gaspari2017}, while extending them here to include self-regulated micro-scale jet feedback across a broader dynamic range. The present runs therefore place the classical CCA condensation problem in a feedback-regulated {\sc BlackHoleWeather} context, where the focus shifts from how multiphase gas forms in a stratified atmosphere to how the ensuing jet reshapes the morphology and thermodynamics of that same atmosphere.

As introduced above, all simulations carried out in this work have been performed with the \athenapk code using a $100$ kpc side box at 10 static refinement levels. Relative to the 12-level setup adopted in B26a, this corresponds to an effective factor $2^3=8$ decrease in spatial resolution, with one additional factor of 2 arising from the doubled box size, and thus to a finest cell size of $\simeq 0.78$ pc. Throughout the whole set of simulations, we assume the same initial density and temperature profiles as for a static hot gaseous halo in hydrostatic equilibrium. A static gravitational potential is computed by considering the contributions of the dark matter halo, the central dominant group galaxy, and a supermassive black hole, whose initial mass of $M_{\rm BH} = 2\times10^{8} M_{\odot}$ is updated accordingly to Eq.~(\ref{eq:bhar}). This value is consistent with a central black hole in a low-redshift galaxy group as derived from X-ray scaling relations \citep{GaspariEckert2019} and with semi-analytic and cosmological models for seeding prescriptions of black hole formation \citep[e.g. ][]{PianaDayal2021, CammelliMonaco2025}.  The initial profiles are discussed in detail in B26a, including the data sample and the fitting procedure used to obtain the initial profiles. Here we note that we opt to simulate a galaxy group (reference parameter values are reported in B26a) as this configuration shows a consistently faster evolution with respect to a galaxy cluster. In fact, the cooling time in a galaxy group, which drives the criterion for the condensation of gas into CCA episodes \citep{Gaspari2013, Gaspari2017}, results in about $t_{\text{cool}}\approx 10-20$ Myr in the central kpc in our simulations, typically one or two orders of magnitude shorter than a galaxy cluster. This allows us to test our implemented physics in a computationally efficient manner and, on the other hand, the large scale structure of the group/cluster is not expected to affect the micro-scale phenomena we have implemented as described in the previous section.

Following the analysis carried out in B26a, in this work, we employ two different macro `weather' scenarios in terms of initial conditions at both high- and low-turbulence to better study SMBH feeding and feedback processes in realistic environments, indicated as \high and \low, respectively. For reference, the main parameters of the turbulence driving used in B26a are shown in Table~\ref{tab:turbulence}. We note that turbulence is driven only through solenoidal modes on a scale of about $\sim25$ kpc, which aligns with values expected from typical AGN feedback stirring and/or sloshing and minor mergers \citep{Gaspari2012, Vazza2012}. Similarly, our reference runs are built according to the following mechanism. For both runs, \high and \low cases, we first let the simulations run in a regime where only the turbulence driving is activated, while keeping radiative cooling and jets off. This allows turbulence to develop to realistic, observed scenarios \citep[for direct and indirect observed velocity dispersion values, see also][B26a,b]{GaspariChurazov2013, Hofmann2016, Xrism2025}. To achieve this, we aim at reproducing two cases that we regard as lower and upper limits by targeting 3D Mach numbers of $\mathcal{M}\sim0.4$ for the \high run and $\mathcal{M}\sim0.15$ for the \low run. In terms of velocity dispersion, these correspond to values of $\sigma_v \simeq 210\ - \ 230~\mathrm{km~s^{-1}}$ and $\sigma_v \simeq 60\ - \ 90~\mathrm{km~s^{-1}}$, respectively, reached approximately after $\sim35$ Myr, time where both the jet injection and radiative cooling are activated.  

However, we note that in this work we investigate a specific case of the injection mechanism introduced in the previous section. In particular, we limit our efforts to studying the effects and consequences of a kinetically-dominated injected jet, i.e., setting the kinetic fraction to unity in Eq.~(\ref{eq:approx_tot_power}). This choice is mainly driven by the particularly interesting, yet vastly unknown mechanism that deposits kinetic energy into other channels at different scales and gas phases, which is the key focus of the results presented in the next section. In addition, we will make use of our findings in future companion papers in the BHW series, in accordance with the project aim to link and study AGN accretion and feedback at different scales and phases as a whole, following via first principles a \textit{bottom-up} paradigm.

\begin{table}
    \centering
    \caption{Turbulence driving parameters.}
    \label{tab:turbulence}
    \begin{tabular}{lccccc}
        \toprule
        \textbf{} & $a_{\mathrm{rms}}$ & $n_{\mathrm{peak,v}}$ & $N_{\mathrm{modes,v}}$ & $\zeta$ & $t_{\mathrm{corr}}$ \\
        & [$10^{-8}\ {\rm cm\,s^{-2}}$] &  &  &  & [Myr] \\
        \midrule
        \texttt{low}  & 0.62 & 4 & 64 & 1 & 30 \\
        \texttt{high} & 1.55 & 4 & 64 & 1 & 30 \\
        \bottomrule
    \end{tabular}
    \tablefoot{
        Parameters of the stochastic turbulence-driving module, with respect to the 2 initial condition runs. The acceleration field has an RMS amplitude $a_{\mathrm{rms}}$ and is injected with a solenoidal fraction $\zeta=1$, equivalent to purely incompressible driving peaking at wavenumber $2\pi\ n_{\mathrm{peak,v}}/L_{\rm box}$. The $N_{\mathrm{modes,v}}$ random Fourier modes are resampled every correlation time $t_{\mathrm{corr}}$. The two sets of simulations only differ in the driving strength $a_{\mathrm{rms}}$. Table adapted from B26a.
    }
\end{table}

\section{Results}\label{sec:results}
In this section, we present the main results of our analysis as the outcome of the \high and \low reference runs introduced in \S\ref{sec:setup}. In particular, our aim is to test how AGN feedback, when injected on micro-scales, affects the surrounding environment in two separate regimes of turbulence driving. 
We define \train as the elapsed time between the activation of cooling and jet feedback and the onset of the first clear precipitation episode, identified by the appearance of cold clumps and filaments together with a corresponding rise in sink accretion. For the two reference runs considered here, \train is approximately 9 Myr for \low and 16 Myr for \high. This increase in sink accretion marks the characteristic onset of the rain phase and therefore sets the natural evolutionary clock of the system. We then use \train to normalize the simulation time and compare the two runs at equivalent evolutionary stages.

With respect to Eq.~(\ref{eq:tot_power}), the parameter $\epsilon$ adjusts the total efficiency of the jet to convert the accreted mass into feedback power. For the sake of this work, we consider the jet geometry to be fixed in time, i.e., the injection region is set to the one described in Figure~\ref{fig:grid} at the maximum refinement level, while keeping the efficiency $\epsilon$ constant. In this way, the accreted mass per unit of time is the only ingredient capable of modulating the jet power. We regard this approach as more stable and representative of the actual physics of injection, although the phenomena that regulate this mechanism are neglected in this work and are parametrized by a single parameter $\epsilon$. Although this approach does not explore micro-scale accretion processes \citep[e.g. ][]{TchekhovskoyNarayan2011}, it allows us to better study and isolate the physical impact of the jet. We envisage investigating the physical scenarios occurring beyond the sink scale in a separate paper of the BHW series (P26a,b). 

Following the multi-scale and multi-phase approach, we separate the gas into four spatial scales, unless stated otherwise: \textit{micro} scale ($r < 0.1\ \mathrm{kpc}$), \textit{meso} scale ($0.1\ \mathrm{kpc} < r < 1\ \mathrm{kpc}$), inner \textit{macro} scale ($1\ \mathrm{kpc} < r < 10\ \mathrm{kpc}$), and outer \textit{macro} scale ($r >10\ \mathrm{kpc}$). We further introduce five thermal phases, corresponding to the temperature intervals listed here: cold molecular gas (`Molecular', $T < 2\times10^{2}\ \mathrm{K}$), cold atomic gas (`Cold', $2\times10^{2}\ \mathrm{K} \leq T < 1.6\times10^{4}\ \mathrm{K}$), warm gas (`Warm', $1.6\times10^{4}\ \mathrm{K} \leq T < 1.16\times10^{6}\ \mathrm{K}$), hot gas emitting in the soft X-ray (`Hot - Soft X', $1.16\times10^{6}\ \mathrm{K} \leq T < 5.8\times10^{6}\ \mathrm{K}$) and hot gas emitting in hard X (`Hot - Hard X', $T >5.8\times10^{6}\ \mathrm{K}$).
We remark that these phase labels are used here as thermodynamic analysis proxies, not as full synthetic observables.

This section is organized as follows. The morphology of the gas is presented in \S\ref{subsec:morpho} by comparing how the gas structure at different scales is shaped by the jet-driven wind in the two different regimes. \S\ref{subsec:stats} inspects how the jet affects the physical state of the gas, focusing on the radial phase profiles and PDFs. The evolution of gas masses in different phases is described in \S\ref{subsec:masses}, while \S\ref{subsec:bhar} presents the accretion rate of the SMBH and the related jet power. Finally, \S\ref{subsec:phase} dissects the phase structure in terms of temporal and radial evolution.

\subsection{Morphology}\label{subsec:morpho}
\begin{figure*}
    \centering
    \includegraphics[width=0.9\linewidth]{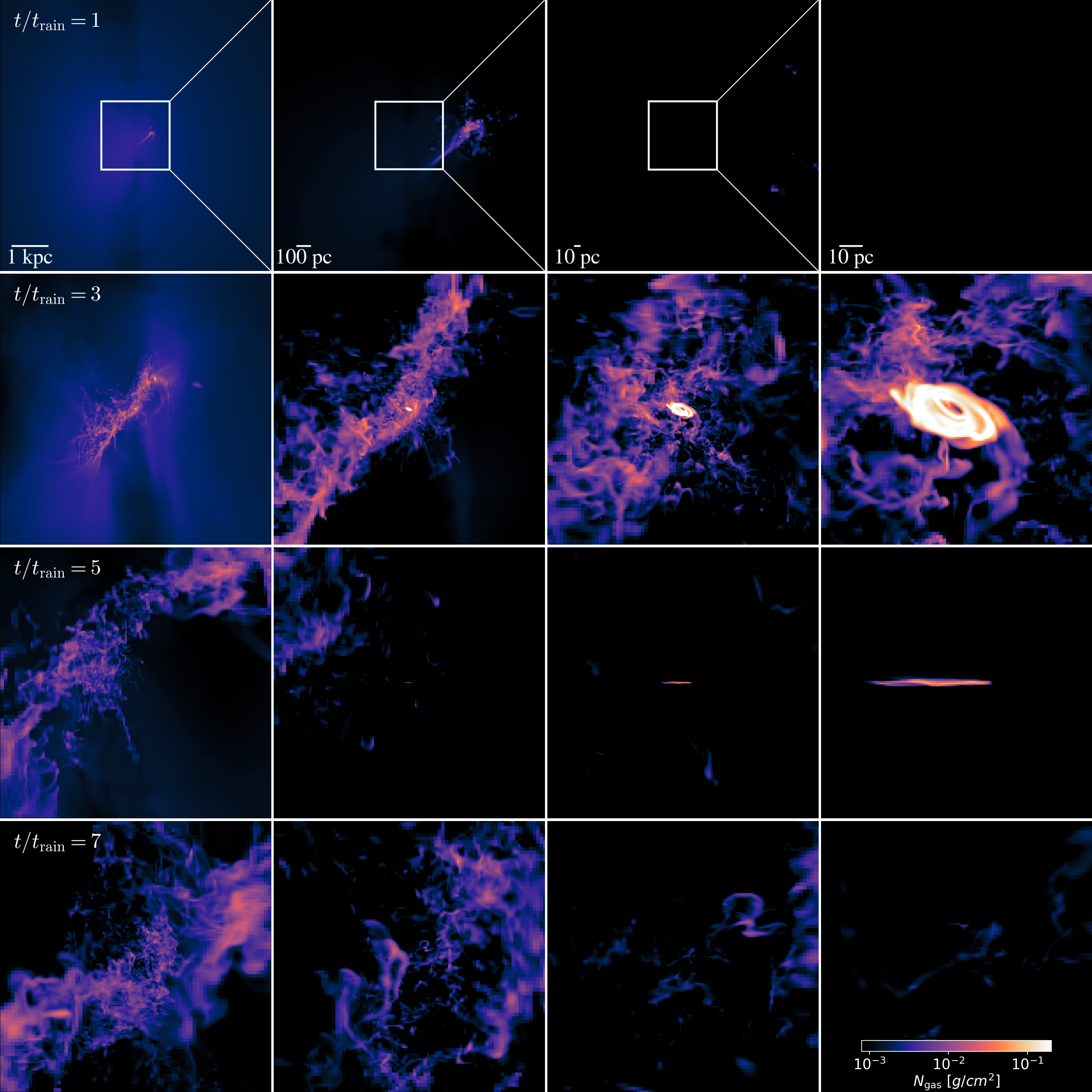}
    \caption{Zoom–cascade of projections of the simulated gas density along the plane perpendicular to the jet injection axis, $z$, in the \high run. Each panel shows a projection through a finite slab whose thickness equals the panel width, centered on the same point; widths decrease along each column, while the time evolution is observed in different rows from top ($t/t_{\rm rain}=1$) to bottom ($t/t_{\rm rain}=7$). In the first row, panels 1–3 include a thin white square outlining the field of view of the subsequent panel to guide the eye. For visual comparability, all panels do share the same colormap, density limits and spatial zooming along each column.}
    \label{fig:mosaic_high}
\end{figure*}
\begin{figure*}
    \centering
    \includegraphics[width=0.9\linewidth]{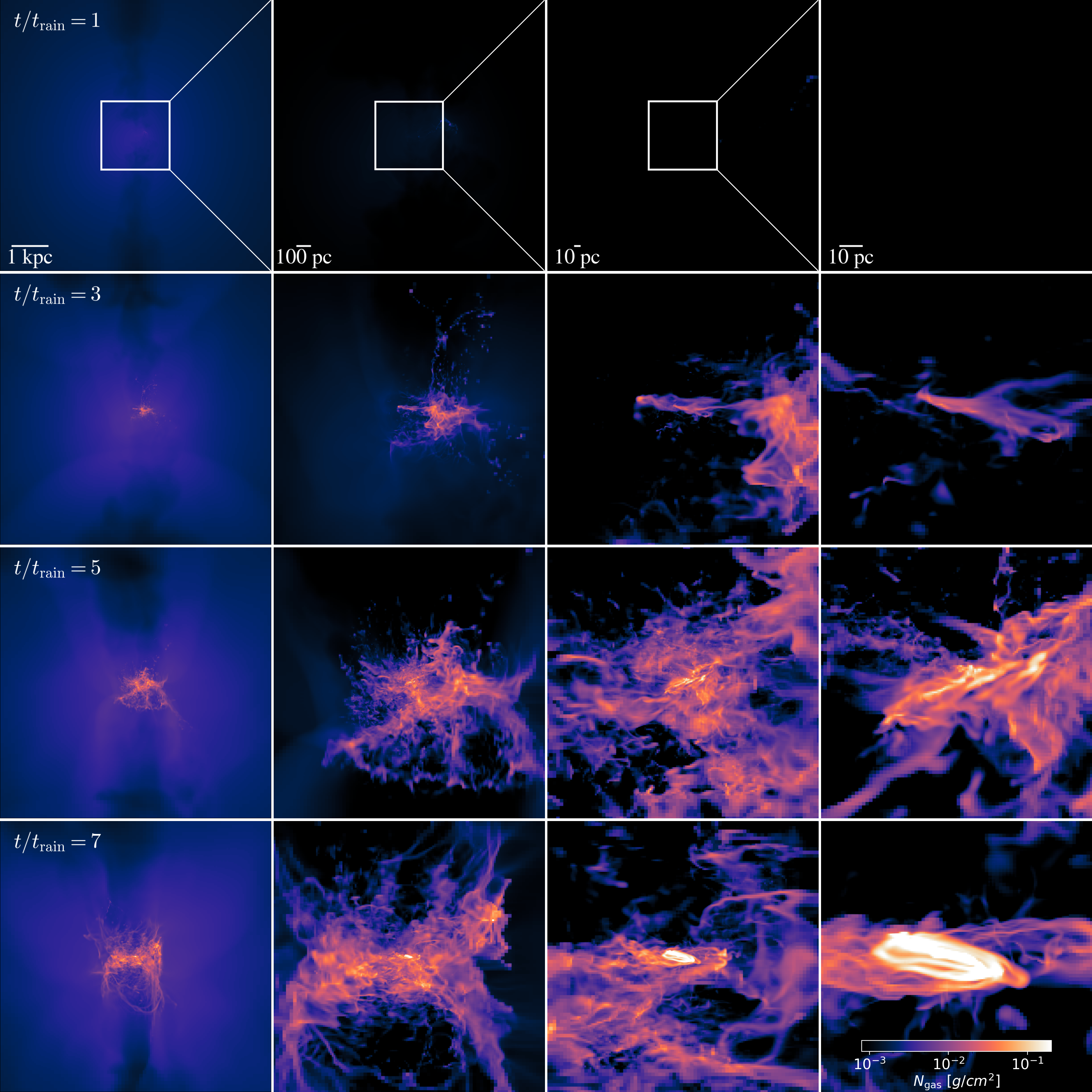}
    \caption{Same as Figure~\ref{fig:mosaic_high}, zoom–cascade of projections of the simulated gas density along the plane perpendicular to the jet injection axis, $z$ in the \low case.}
    \label{fig:mosaic_low}
\end{figure*}
\begin{figure*}
    \centering
    \includegraphics[width=0.9\linewidth]{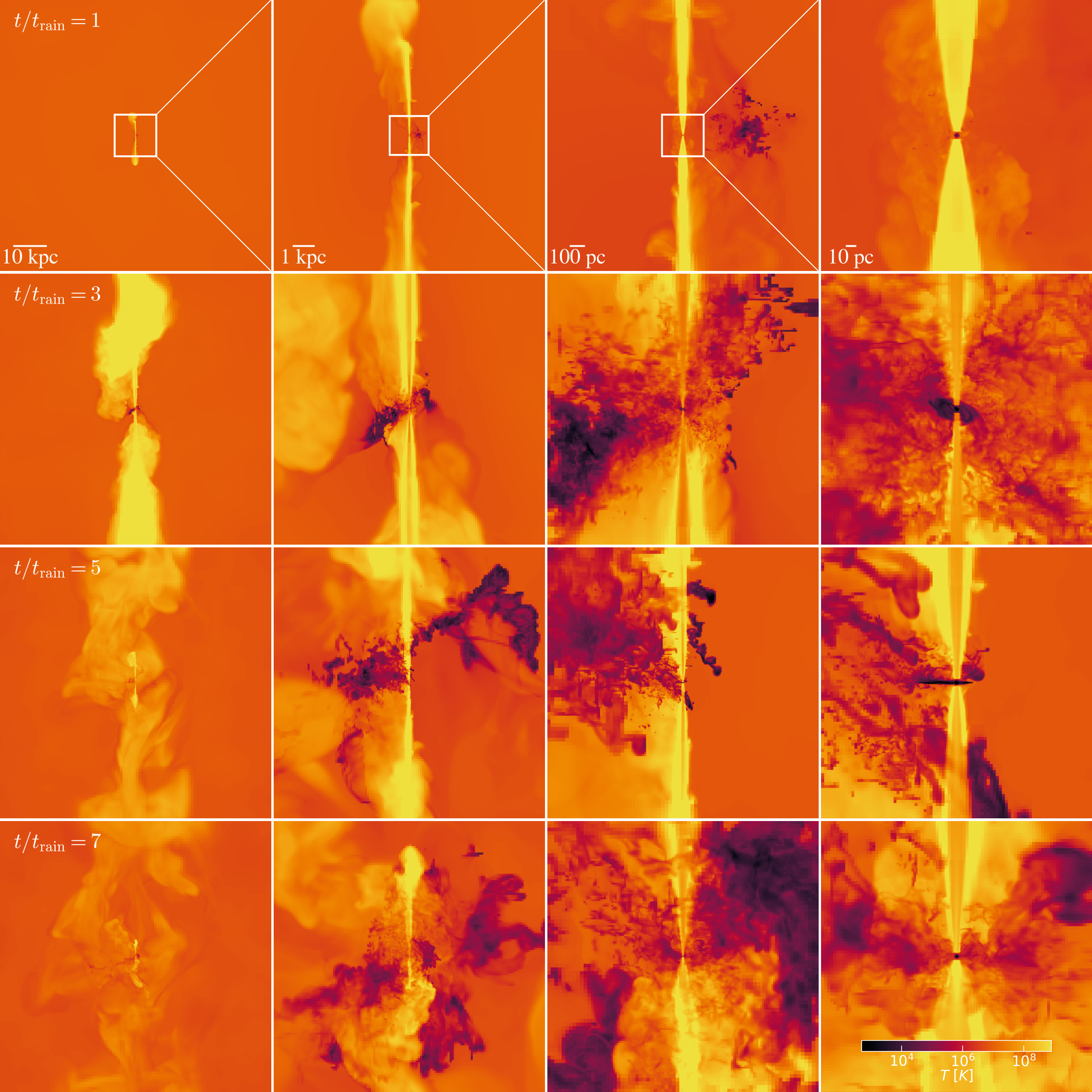}
    \caption{Similar to Figure~\ref{fig:mosaic_high}, zoom–cascade of projections of the gas temperature along the plane perpendicular to the jet injection axis, $z$, in the \high run. In this case, to show the spatial expansion of the jet material and entrained/heated gas, the projected volumes start from approximately the full box, zooming down to the pc scale. For visual comparability, all panels do share the same colormap and temperature limits. The time evolution shows that the cocoon becomes broader, more asymmetric, and more porous, producing a wider and more irregular jet-ambient interface.}
    \label{fig:mosaic_high_temp}
\end{figure*}
\begin{figure*}
    \centering
    \includegraphics[width=0.9\linewidth]{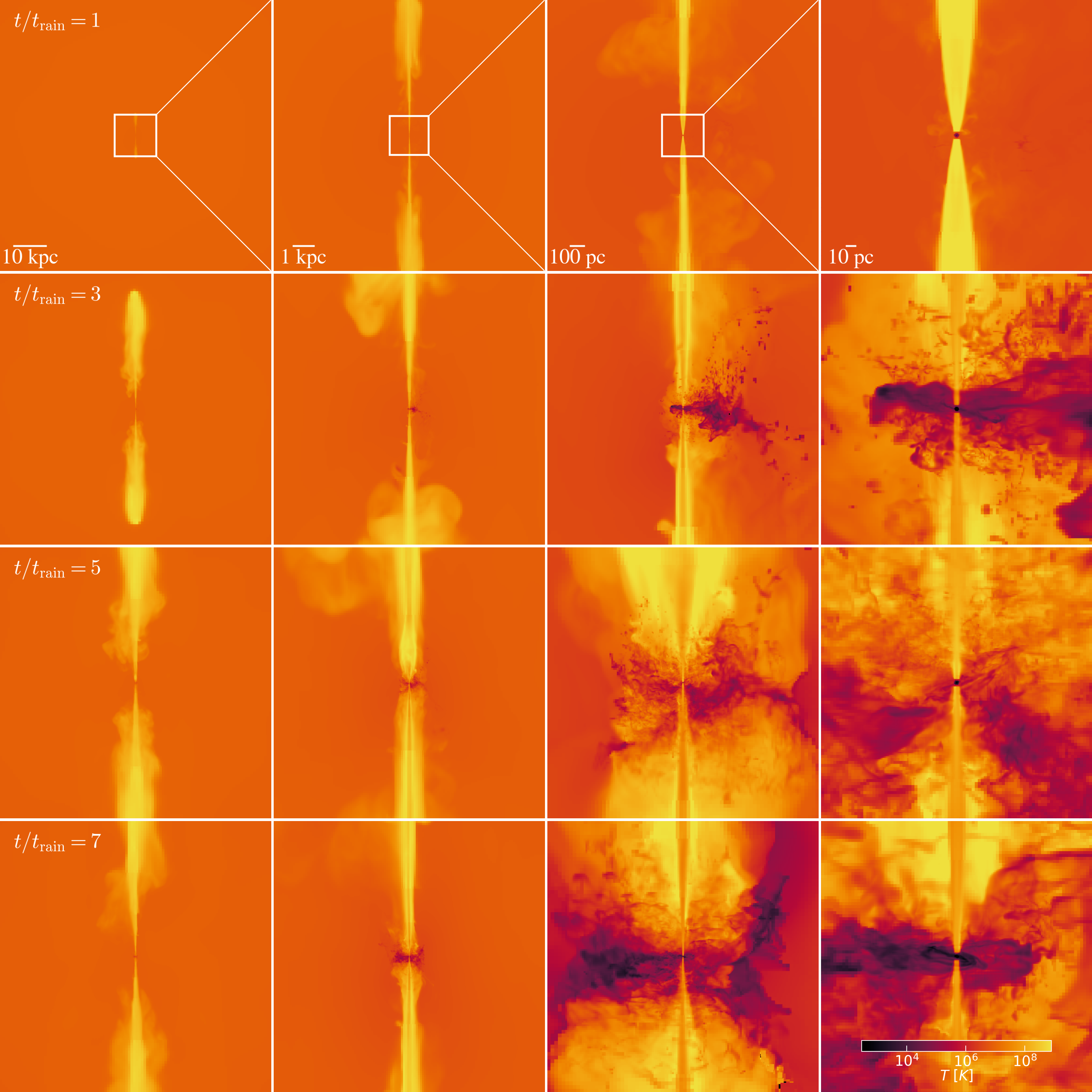}
    \caption{Same as Figure~\ref{fig:mosaic_high_temp}, zoom–cascade of projections of the simulated gas temperature along the plane perpendicular to the jet injection axis, $z$ in the \low case. Over time and scales, the jet remains comparatively collimated and inflates a more coherent bipolar hot cocoon.}
    \label{fig:mosaic_low_temp}
\end{figure*}

Over time, non linear thermal instabilities grow and develop due to turbulence cascade, resulting in chaotic motions. Fluctuations in density and temperature perturb the hot gas and cause condensation in a multiphase medium \citep{Gaspari2013,Gaspari2017}. Hence, the magnitude of the turbulence driving (see \S\ref{sec:setup}) strongly influences the process in which the hot homogeneous gas first cools down, fragments, and eventually precipitates onto the central SMBH in a sort of \textit{rain} analog. In this picture, the \high and \low reference runs studied below represent two idealized experiments to test different environments, without necessarily describing a fully realistic environment of galaxy groups and clusters. In fact, the actual initial conditions depend on a much broader range of ingredients, such as mergers, sloshing, and feedback phenomena. In this section, we emphasize how the turbulence driving affects the subsequent evolution of the system and how its combined action with the jet shapes the morphology.

To visualize the multi-scale and multi-phase structure, Figures~\ref{fig:mosaic_high} and \ref{fig:mosaic_low}, we render a $4\times4$ “zoom–cascade” of line–of–sight projections of the gas density towards the central region, perpendicular to the jet axis, $z$ (from south to north in the projected plane). The $4\times4$ layout is designed to connect kpc-scale jet/atmosphere coupling to the formation of pc-scale cold filaments, clouds, and central dense structures. For each panel, we compute a projection along the plane perpendicular to the jet injection axis through a finite slab whose thickness equals the projected width, centered at the sink particle location. The “thickness$=$width” choice preserves an approximately square sampling volume on each scale. Projection sizes decrease from left to right, as shown by the white squares and the annotated scales in order to aid visual tracking. Instead, the time evolution is followed from the top to the bottom row, at $t/t_{\rm rain}=1,3,5,7$. For visual clarity, we opt to use the same color bar limits across different panels. 

\subsubsection{The \high run: fragmented, extended multiphase development}
Figure~\ref{fig:mosaic_high} shows the \high run. We note here that the rows compare equivalent evolutionary stages in normalized time, not in the same absolute elapsed time. In general, the jet deeply affects the gas structure along the $z$ axis, while the region outside the two cylindrical jet injection areas remains relatively unperturbed. At large scales ($\gtrsim 1$~kpc), early bipolar depressions, marked by the bi-conical low density region, are seen with dark depressed along the $z$ axis, consistent with the line–of–sight projection of jet–inflated cavities bounded by shocks and shear layers. At these scales, the piercing power of the injected material is particularly evident. In the intermediate regime ($\sim 10^2$~pc), the projected morphology is dominated by off-center filaments and clumpy substructure arising from turbulent-driven cooling due to local thermal instabilities, leading to cold gas condensation. Within a few pc, the circumnuclear region exhibits strongly inhomogeneous features, displaying a huge variety of environments, both in terms of spatial distribution and gas phases: cold gas appears in spiral inflows, filaments, and compact condensed clumps, yet coexisting with hotter jet plasma. 

In terms of time evolution, at the early stages ($t/t_{\rm rain}=1$), the high-turbulence medium already exhibits appreciable asymmetry and small-scale structure in the central region, such that the jet propagates into an inhomogeneous background due to the initial turbulence driving. Even at intermediate zoom levels, weakly condensed features appear as patchy overdensities rather than a smooth, axisymmetric configuration. By $t/t_{\rm rain}=3$, the cold phase expands into a complex morphological network: filaments and clumps emerge in multiple directions, and the structures are highly irregular, consistent with strong turbulent stirring and intermittent compression. The jet--atmosphere interaction region is correspondingly distorted, with cold material distributed in fragmented strands rather than confined to a single coherent layer.

By the time the rain has fully developed ($t/t_{\rm rain}\gtrsim3$), the filamentary component becomes more extended and dynamically intricate: elongated features appear kinked and braided, with numerous sub-filaments and knots down to the parsec scale. A compact dense feature develops close to the sink at the deepest zoom, but remains embedded in a turbulent, multiphase environment rather than forming in isolation. As discussed also in B26a, this rotational structure mainly composed by cold clumps stands out especially between $t/t_{\rm rain}=3$ and 5, with typical radius of 10 pc, likely due to extra angular momentum as a result of residual turbulent eddies in the hot gas \citep[see also ][]{Gaspari2015}. Given the lack of central resolution and our modeling of sink particle ($r_{\rm sink}\simeq 3.12$ pc), we do not interpret this as a fully resolved accretion but rather as a clumpy torus like structure, similar to what is observed in the circumnuclear region of AGNs. At late times ($t/t_{\rm rain}=7$), the central region enters a strong turbulent region, generated from the initial driving, which, in turn, dissolves and drags the majority of the cooling/condensing gas. 
In the present setup this late-time stirring is externally driven by construction, although it may qualitatively mimic unresolved large-scale drivers such as mergers, sloshing, or broader AGN-induced turbulence.

In general, the \high case is strongly multiphase across all zoom levels. The morphology is characterized by (i) a broad and chaotic distribution of cold gas on $\sim$100~pc scales, (ii) persistent fragmentation into clouds and filament segments, and (iii) a dense central structure that coexists with continued inflow/outflow and mixing. The \high case produces more spatially widespread cold condensation, with the jet amplifying fragmentation and promoting a filament-dominated appearance up to $\lesssim10$ kpc scale.

\subsubsection{The \low run: more coherent condensation and central settling}
In the low-turbulence run, the early atmosphere remains relatively smooth and symmetric
at $t/t_{\rm rain}=1$, with minimal small-scale cold structure at any zoom level. The jet
therefore interacts with a more uniform medium, and the cold phase is initially weak and limited to the inner $\sim100$ pc range. At $t/t_{\rm train}\gtrsim3$, condensed material appears only in limited regions and remains relatively compact: the morphology is dominated by fewer structures, with less evidence for a tangled, multi-directional filament network. This contrasts with the \high case, where numerous sites of condensation are present in comparable evolutionary stages and extend over a much larger region.

At $t/t_{\rm rain}=3$, condensation becomes more prominent but retains a spatially organized distribution, with fewer major filaments and a reduced level of small-scale fragmentation. The deepest zoom reveals the emergence of a compact, dense feature near the sink. By $t/t_{\rm rain}=5$, the cold phase is strongly concentrated toward the center, although reaching distances of a few hundred pc, while forming a bright and compact structure indicating efficient settling and/or rotational support at small radii. Despite the fact that filaments and sheared features are still visible, they are less spatially pervasive and less topologically complex than in the high-turbulence run. In short, lower ambient turbulence advances the onset of runaway cooling and favors a morphology in which the cold gas is less fragmented and more centrally concentrated at late times.

\subsubsection{Temperature maps}
%
In order to further support the information depicted by the density projections in Figures~\ref{fig:mosaic_high} and ~\ref{fig:mosaic_low}, we also show, at the same times, the respective projections in average temperature in Figures~\ref{fig:mosaic_high_temp} (\high) and \ref{fig:mosaic_low_temp} (\low). This perspective emphasizes from a morphological point of view the jet imprints of CCA rain scenarios on the local gas structure. In fact, this is reflected by the multiphase, turbulent core that mediates energy and momentum coupling between the jet and the ambient medium. As before, rows trace the temporal evolution ($t/t_{\rm rain}=1,3,5,7$ from top to bottom), while columns zoom progressively from the halo scale down to the central $\sim$pc-scale launching region. These maps suggest that the hot phase morphology is associated with the AGN feedback cycle and provide the thermal, complementary view to the filament/cloud structures seen in the density projections.

At early times, both runs display a narrow bipolar jet with a relatively compact hot cocoon. As the system evolves, the jet inflates a broader hot structure and its interaction with the ambient medium becomes increasingly sensitive to the turbulent state of the atmosphere. In the \low run, the jet remains comparatively well collimated and the cocoon retains a more coherent bipolar morphology. In the \high run, instead, the hot structure becomes broader, more asymmetric, and more corrugated, with a wider jet-ambient interface and stronger lateral stirring. These differences closely recall the behavior seen in the density morphology. A more coherent hot channel in the low-turbulence case is associated with a more centrally concentrated condensation, while the broader and more irregular cocoon in the high-turbulence case is consistent with more extended and more fragmented precipitation. The temperature projections therefore make clear that the condensed phase develops within a hot feedback flow whose geometry is itself controlled by the atmospheric weather and potentially supplements further condensation due to the jet turbulence cascade. In the following sections, we provide a more quantitative analysis of the physical phenomena qualitatively described above. 

\subsection{Multi-scale and -phase analysis}\label{subsec:stats}
\subsubsection{Phase radial profiles}
\begin{figure*}
    \centering
    \includegraphics[width=1\linewidth]{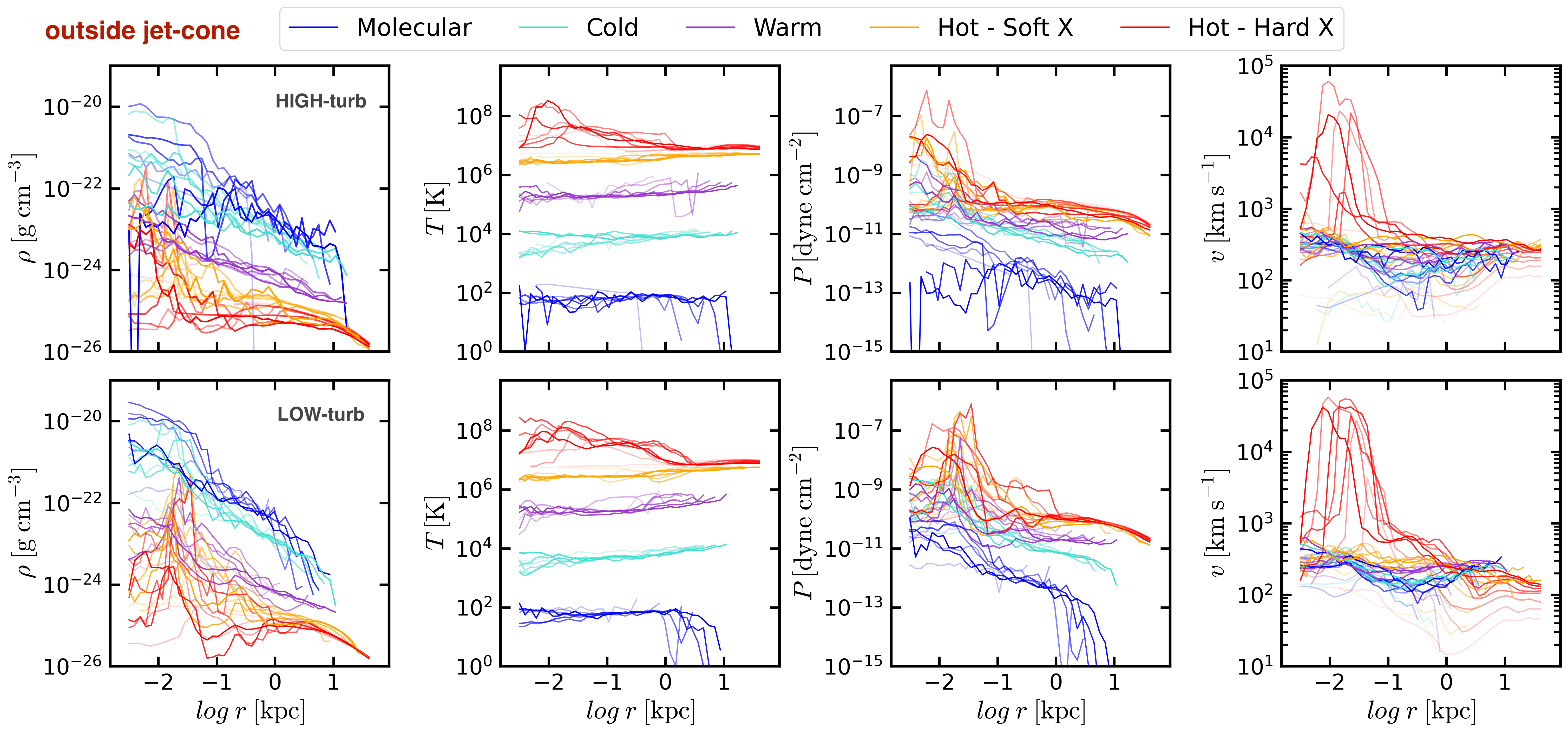}
    \includegraphics[width=1\linewidth]{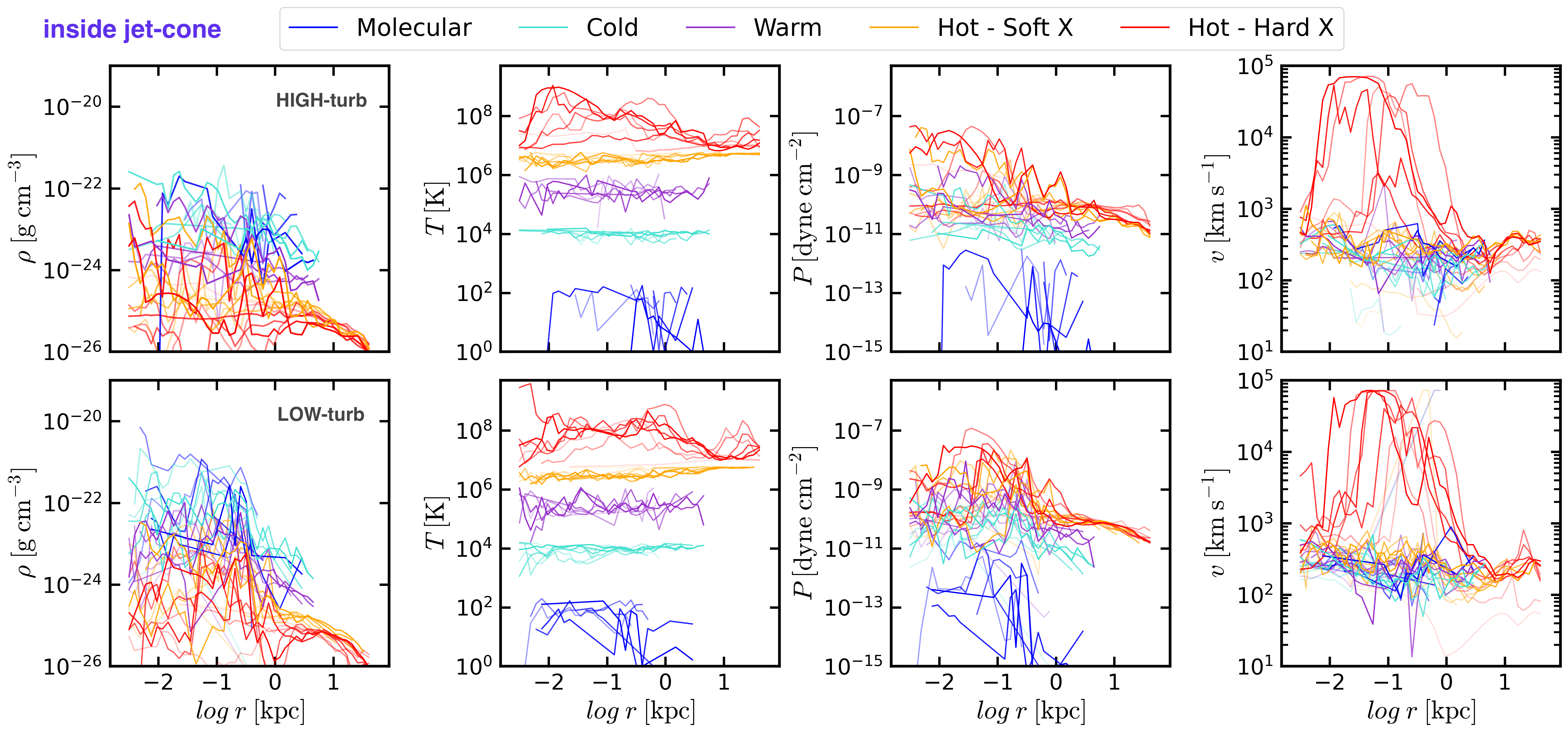}
    \caption{Time evolution of mass-weighted radial profiles of the region outside/inside the jet-cone in the upper/lower figures, respectively. From left to right, several quantities are shown; gas density, temperature, pressure and velocity magnitude with respect to the \high (top rows) and \low (bottom rows) runs. Colors denote thermal phases broadly tied to observational bands (radio, optical, UV, X-rays), i.e., gas selected by broad temperature intervals defined in the text. Darker colors indicate later evolutionary stages. Profiles are calculated in spherical shells, where the radius is measured from the origin, where the sink and SMBH are located. The radial trends highlight how jet-regulated CCA separates the hot, rarefied intra-cone channel from the extra-cone multiphase atmosphere, while the two turbulence regimes differ mainly in the radial extent and scatter of the condensed phases.}
    \label{fig:profiles_evo}
\end{figure*}

Figure~\ref{fig:profiles_evo} presents radial, phase-separated profiles of the thermodynamic and kinematic state of the gas (gas density, temperature, pressure and velocity magnitude), computed separately (i) outside (\emph{extra-}, top panel) the jet cone, defined as the volume enclosed within a 15$^\circ$ half-opening angle aperture around the $z$-axis, and (ii) inside (\emph{intra-}, bottom panel) the jet cone (the latter excludes the jet spine to suppress contamination from extremely low-mass cells in the jet channel, defined as a 2$^\circ$ half-opening angle cone). The upper rows depict the \high case, while the \low scenario is instead shown in the bottom rows. The color gradient represents the time evolution of the profiles, with lighter/darker gradients indicating earlier/later times, respectively. In the extra-cone region the atmosphere develops a clear multiphase stratification: at fixed radius the molecular and cold components occupy the highest densities, the warm phase lies at intermediate densities, and the soft/hard X-ray phases trace the most rarefied material. However, at radii $\lesssim 0.1$ kpc, the stronger time and spatial variability of the phase profiles reflect the sample variance introduced by considering smaller and smaller volumes. The temperatures of each phase remain close to their defining ranges with only mild radial trends, while the pressure profiles exhibit substantial overlap among phases at smaller radii ($\lesssim 0.1$ kpc), consistent with approximate (though not perfect) pressure balance between condensed structures and the surrounding hot medium; as well as for the density curves, deviations and increased scatter toward small radii reflect the strongly time-dependent jet-IGrM interaction and turbulent stirring. The velocity profiles further separate the phases dynamically: the cold/molecular gas typically remains in the few $\times 10^{2}\ \mathrm{km\ s^{-1}}$ range, whereas the X-ray emitting plasma shows systematically higher velocities and intermittent high-speed excursions associated with jet-driven motions and entrainment propagating outside the jet cone. Inside the jet cone the profiles change qualitatively, highlighting the formation of a low-density, hot, dynamically active channel: the soft/hard X-ray phases dominate the cone statistics, with elevated temperatures and higher characteristic velocities, while the cold and molecular phases are strongly suppressed and only appear intermittently over a restricted radial range, indicative of transient mixing/entrainment at the jet-ambient interface rather than sustained in-situ condensation within the channel. 

Comparing the two turbulence regimes (top versus bottom rows), the extra-cone region highlights that the presence of the jet has the effect of alleviating the differences between the \high and \low case runs. In fact, temperature, pressure and velocity magnitude exhibit similar trends, where minimal deviations can be regarded as due to statistical fluctuations and sample variance, and the overall behavior is rather consistent. Nonetheless, it is important to note that the \low density profiles extend up to a slightly smaller ($\sim 2/3$ times) radius with respect to the \high case, although the projected frames shown in Figures~\ref{fig:mosaic_high} and \ref{fig:mosaic_low} might induce a more extreme interpretation. This effect is a consequence of the jet and is independent on the turbulence driving regime: the extra turbulence support introduces a new channel for the formation of cold filaments condensing along the ridges of the Kelvin-Helmholtz curls left by the interaction between the jet and the ambient medium, where the typical time scale of Kelvin-Helmholtz $t_{\rm KH}\sim 10^{5-7}$ yr is typically shorter than the cooling time $t_{\rm cool}\sim10-20$ Myr (see B26a), triggering local instabilities ($t_{\rm cool}/t_{\rm KH}\gtrsim1$). The net result dilutes the initial imprint of the driven turbulence, in contrast to what is shown in B26a where the two cases have distinct trends. In the intra-cone region, the \high case exhibits broader scatter and stronger radial overlap between phases, consistent with the picture of a more efficient mixing and a more porous jet channel, whereas the \low case maintains cleaner phase separation and a cone that is more uniformly hot and rarefied.

A useful comparison can be made with B26a, which isolates the role of large-scale turbulent stirring in shaping the multiphase condensation cascade. In those simulations, the molecular component reached central densities of $\sim10^{-19}$--$10^{-18}\,{\rm g\,cm^{-3}}$, particularly within the inner $\sim100$ pc. In the present jet-regulated runs, the molecular densities are lower by about two orders of magnitude. While B26a shows how turbulence alone can promote the formation of a dense, centrally concentrated molecular reservoir, the jet-regulated case introduces uplift, entrainment, and jet--ambient mixing, which redistribute part of the condensed gas along cavity edges and interface regions, making the turbulence driving support less distinguishable. This suggests that explicit jet feedback does not erase the turbulence-regulated condensation pathway, but modifies it by lowering the peak molecular density and broadening the spatial distribution of the cold phase.

\subsubsection{Probability density functions}
\begin{figure*}
    \centering
    \includegraphics[width=0.9\linewidth]{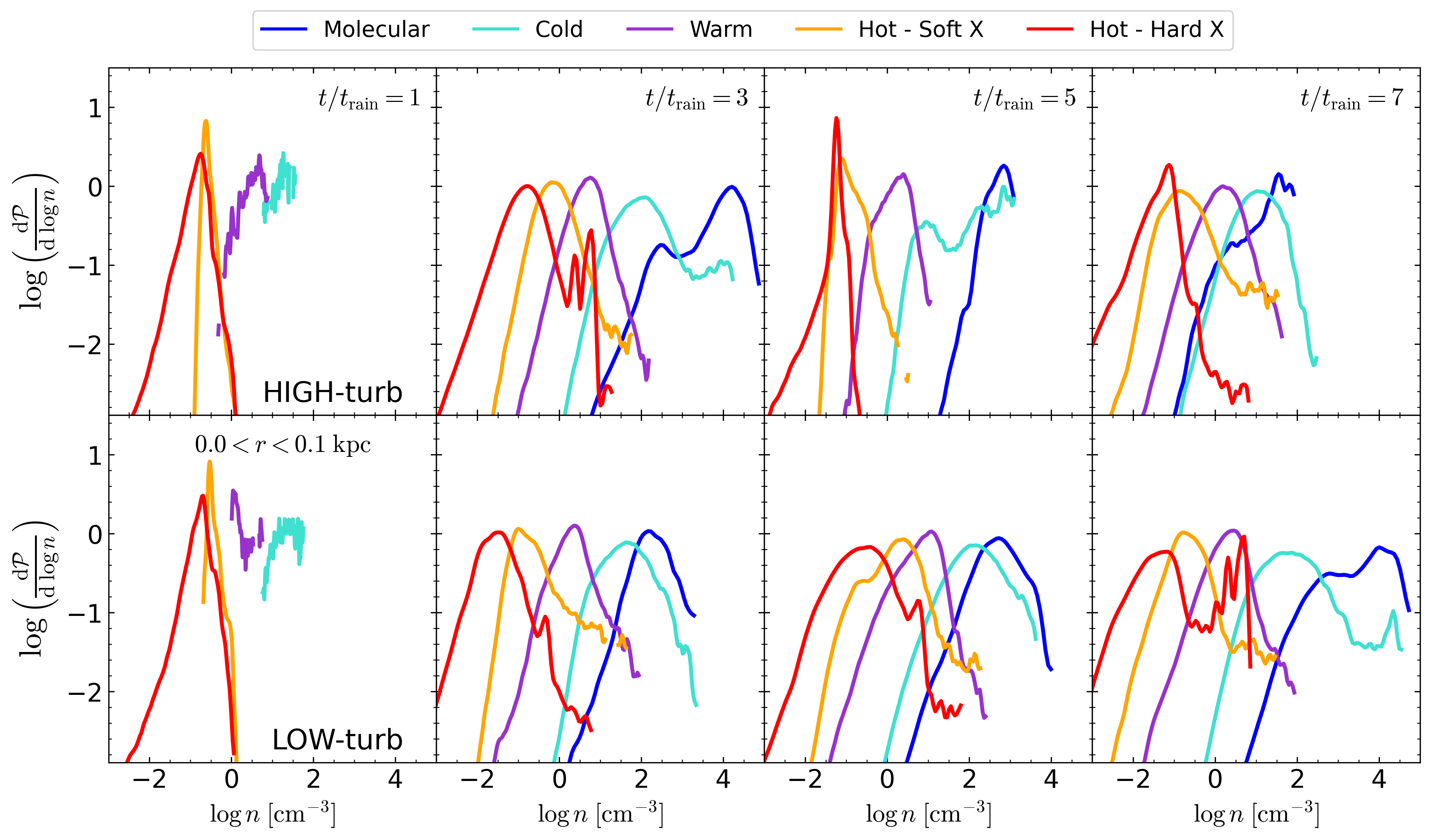}
    \caption{Number–density logarithmic probability distribution functions (PDFs) in the radial shell $0<r<0.10~\mathrm{kpc}$ (micro-scale) are shown at four times ($t/t_{\rm rain}=1,3,5,7$) with respect to the \high (top row) and \low (bottom row) case runs. Colors denote gas phases (Molecular, Cold, Warm, Hot - Soft X, Hot - Hard X; see legend), i.e. gas selected by broad temperature intervals as defined in the text. In the innermost shell, the early distribution, $t/t_{\rm rain}=1$, of both scenarios is dominated by hot/rarefied gas (X–ray colors), while a few gas clouds have begun to cool down to the warm and cold phases. 
    With time ($t/t_{\rm rain}\gtrsim3$), both runs develop broad high-density tails as cool gas condenses near the sink. The \high\ run reaches its largest cold/molecular densities around $t/t_{\rm rain}\sim3$ and then shifts toward lower characteristic densities, consistent with a transition from \stormy\ to \cloudy\ behavior. The \low\ run instead retains a denser and more persistent inner cold component at late times.}
    \label{fig:pdf_phases_micro}
\end{figure*}

\begin{figure*}
    \centering
    \includegraphics[width=0.9\linewidth]{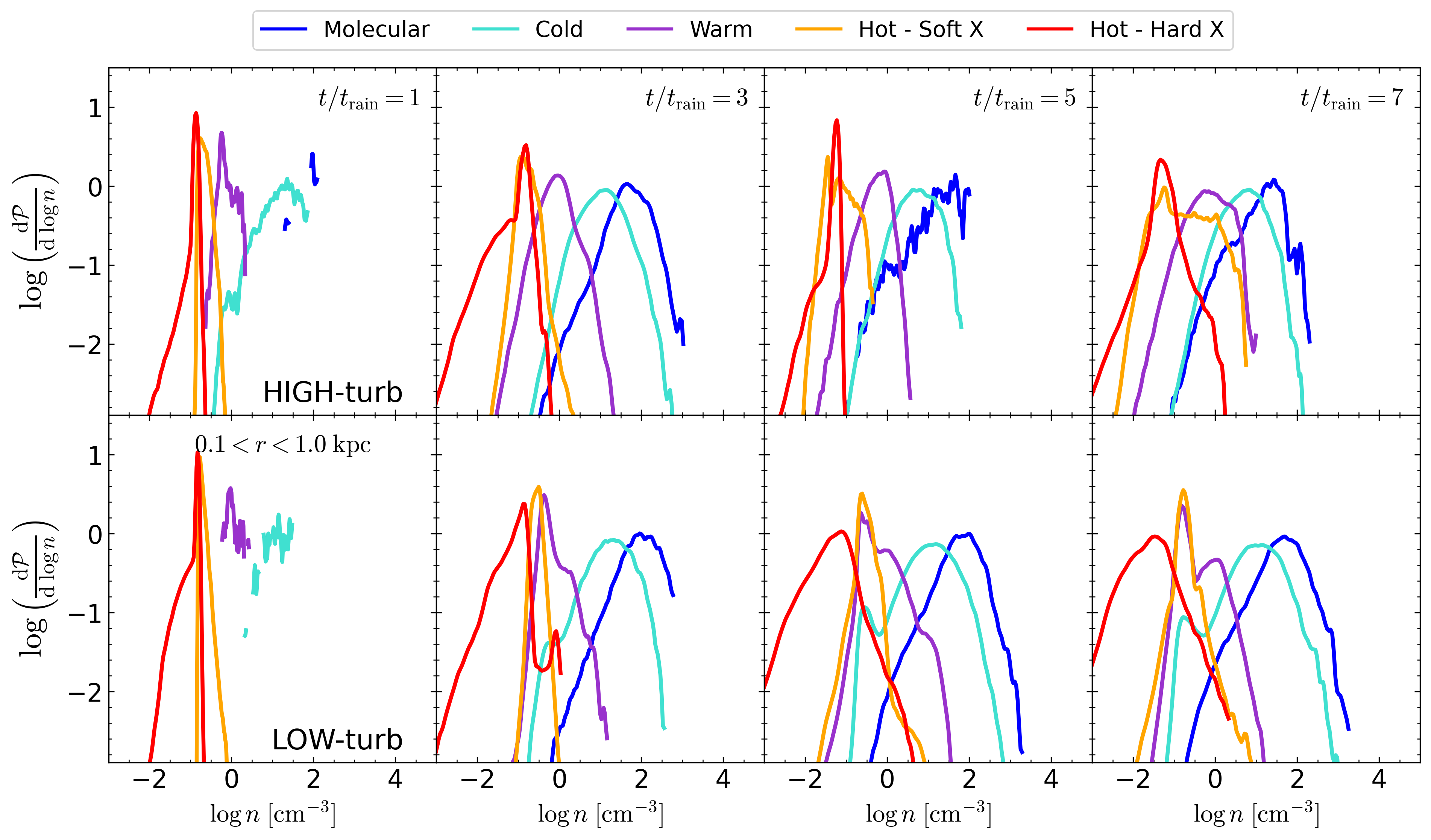}
    \caption{Same as Figure~\ref{fig:pdf_phases_micro} in the radial shell $0.10<r<1.0~\mathrm{kpc}$. At meso scales, the PDFs evolve from a narrow, hot distribution to a multiphase state with still substantial spread in density, although slightly depressed at the $n\gtrsim10^{3}$ especially in the \high scenario. The emergence of cooler bands at later times signals local thermal instability and uplift–assisted condensation in the inner kpc, even though in a more limited amount with respect to the micro scale (Figure~\ref{fig:pdf_phases_micro}).}
    \label{fig:pdf_phases_meso}
\end{figure*}
\begin{figure*}
    \centering
    \includegraphics[width=0.9\linewidth]{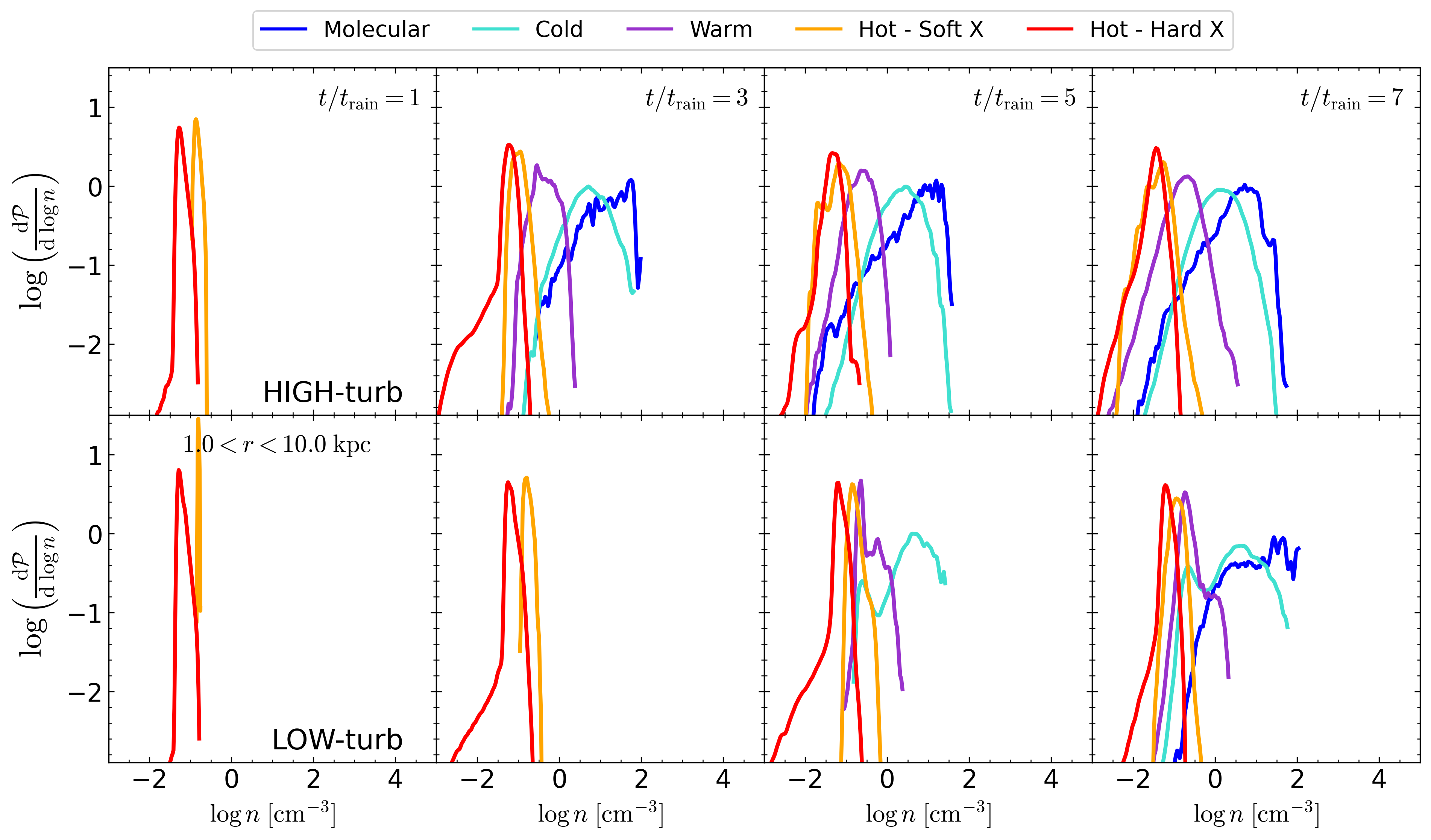}
    \caption{Same as Figure~\ref{fig:pdf_phases_micro} in the radial shell $1.0<r<10.0~\mathrm{kpc}$. In the inner macro region, the distributions remain dominated by X–ray hot, low–density gas for most of the evolution for both reference runs, with the key difference that the \high case displays extended cooled gas, starting from $t/t_{rain}\gtrsim3$ and spanning throughout the whole evolution, showing the more spatially extended rain occurring in this regime. }
    \label{fig:pdf_phases_macro}
\end{figure*}

To quantify how the multiphase medium builds up across scales, we compute the (normalized) mass weighted number-density probability density functions (PDFs), $\mathrm{d\mathcal{P}}/\mathrm{d}\log n$, separately for each thermal phase and in three radial bins. Following the approach of B26a, we use mass-weighted PDFs. The mass fraction in each density bin $i$ is defined as $p_i \equiv \frac{1}{M_{\rm shell}}\sum_{j\in i} m_j$, where $m_j$ is the mass of cell $j$ and $M_{\rm shell}=\sum_j m_j$ is the total gas mass in the shell. Hence, the differential PDF is $\mathrm{d\mathcal{P}}/\mathrm{d}\log n \simeq p_i/\Delta\log n$, with $\Delta\log n$ the logarithmic bin width. By construction, $\sum_i p_i = 1$ such that the integral over a given $\log n$ interval results in the corresponding mass fraction. We use 150 bins in the range $\log n=-3$ to 6 and for each phase we calculate the mass-weighted mean $\mu_{\log n}$ and the standard deviation $\sigma_{\log n}$ and the skewness $\mathcal{S}_{\log n}$. The mean measures the characteristic density of each phase, the dispersion traces the spread of local conditions, and the skewness indicates the relative importance of the high- or low-density tails. 

Figures~\ref{fig:pdf_phases_micro}, \ref{fig:pdf_phases_meso} and \ref{fig:pdf_phases_macro} summarize the time evolution of the number density PDFs in three concentric shells that highlight the micro-, meso- and inner macro scales (outer-scale are not shown since only the hot, volume filling phase, is present and no conclusions can be drawn in terms of multiphase gas). At all radii and epochs, the hot X-ray emitting components (soft/hard-X; orange/red) occupy the lowest densities and retain comparatively narrow peaks, whereas progressively cooler phases shift to higher densities, with the warm phase (purple) bridging toward the cold (cyan) and molecular (blue) components. The emergence and broadening of the high-density tail therefore provide a direct statistical signature of condensation and fragmentation. Overall, the PDFs reveal a radial stratification: condensation is strongest at small radii, while the outer halo stays hot and dilute, consistent with self-regulated cooling, as predicted by CCA, and mechanical feedback. This radial trend highlights that sustained cooling and dense phase growth are concentrated toward the center, while the halo largely remains hot and dilute. The phase-resolved PDFs reinforce the morphological picture from the multiscale maps and corroborate the results of B26a showing the persistence of CCA rain regardless of the presence of the jet, depicting a more extended and chaotic precipitation (\stormy weather) in the \high run, while a more compact and sustained rain (\rainy weather) in the \low scenario (weather states are discussed in more depth in \S\ref{sec:weather_sequence}). In fact, the higher ambient turbulence leads to more complex and radially extended condensation, producing broader and more mixed density distributions, while lower turbulence advances precipitation and keeps the multiphase component more centrally concentrated. However, a crucial difference emerges from the present analysis when looking at the radial dependence of the PDFs.

In the nucleus ($r<0.1$ kpc, Figure~\ref{fig:pdf_phases_micro}), the distributions are initially narrow and hot, but progressively broaden and develop a pronounced high–density tail; cool components grow with time, consistent with precipitation and inflow toward the center. Both turbulence regimes evolve from a primarily hot component at $t/t_{\rm rain}=1$ toward a clearly multiphase density distribution at later times.  By $t/t_{\rm rain}\sim3$, the condensed phases populate densities of orders of magnitude above the hot peak, and the PDFs become broad and partially overlapping, consistent with a strongly mixed environment in the immediate vicinity of the sink. The \high run exhibits molecular and cold phases reaching their largest mean densities around $t/t_{\rm rain}\sim3$ ($\mu_{\log n, \rm mol}\simeq3.82$ dex and $\mu_{\log n, \rm cold}\simeq2.15$ dex), but then shift downward by late times, to $\simeq1.32$ and $\simeq1.04$ dex at $t/t_{\rm rain}=7$ in the central $\sim$100~pc. The $t/t_{\rm rain}\sim3$ epoch marks the peak of the density reached by the \high run, where the cold and molecular components reach their largest characteristic densities up to $n \sim 10^5 \ \mathrm{cm}^{-3}$. At later times ($t/t_{\rm rain}=5$--7), however, the inner PDFs show that the cold and molecular peaks shift toward lower densities, rather than continuing to move to higher $n$. At the same time, the molecular skewness becomes strongly negative ($\mathcal{S}_{\log n, \rm mol}\simeq-1.33$), indicating a growing low-density tail. This quantitatively supports the interpretation that the early inner precipitation storm gives way to a \cloudy phase in which dense gas remains present but is increasingly mixed, redistributed, or accreted rather than continuously building up into the most extreme overdensities. The \low run behaves differently: the molecular and cold means remain higher at late times ($\mu_{\log n, \rm mol}\simeq3.52$ dex and $\mu_{\log n, \rm cold}\simeq1.94$ dex at $t/t_{\rm rain}=7$), consistent with a denser and more persistent central cold reservoir. Additionally, the cold/molecular PDFs continue to extend towards larger $n$. This suggests that the central multiphase medium of the \high run has entered a different regime, in which dense condensates are no longer building up monotonically but are instead being mixed and/or disrupted. We interpret this evolution as the transition from an early \stormy phase, dominated by vigorous precipitation and rapid density growth, to a later \textit{cloudy} phase, in which the inner shell still hosts a substantial cold/molecular component but in the form of a more dispersed population of dense clouds rather than an extreme high-density precipitation core.  The \low run follows the same overall sequence more gradually, with a more persistent dense component, consistent with its sustained and more centrally retained condensation cycle typical of the \rainy phase (see also B26a).

In the meso-scale shell ($0.1<r<1$~kpc) in Figure~\ref{fig:pdf_phases_meso}, the same qualitative evolution is present, although in a shallower form than in the nucleus. By $t/t_{\rm rain}\sim3$, both runs have developed well-defined warm, cold, and molecular PDFs, with the \high case again showing the mean densities of the molecular and cold phases subsequently decline from $\mu_{\log n, \rm mol}\simeq1.71\rightarrow1.30\rightarrow1.18$ dex and $\mu_{\log n, \rm cold}\simeq1.06\rightarrow0.73\rightarrow0.71$ dex between $t/t_{\rm rain}=3$, 5, and 7, while the molecular skewness becomes strongly negative, reaching $\mathcal{S}_{\log n, \rm mol}\simeq-1.14$ at $t/t_{\rm rain}=5$ and $\simeq-0.95$ at $t/t_{\rm rain}=7$. This behavior is weaker than in the inner shell, but points in the same direction: the meso-scale atmosphere is no longer dominated by the rapid density growth of a precipitation storm, and instead transitions toward a \cloudy regime in which condensed structures remain abundant but are more distributed in density space and less dominated by the most extreme overdensities. 
On the other hand, the \low run shows a smoother late-time evolution in the same shell, with the molecular mean density declining more mildly from $\mu_{\log n,\rm mol}\simeq 1.96 \to 1.80 \to 1.64$ dex between $t/t_{\rm rain}=3,5,7$, while the cold phase evolves from $\mu_{\log n,\rm cold}\simeq 1.25 \to 1.03 \to 0.98$ dex. Thus, although both runs show some late-time redistribution of the cool gas at meso scales, the \low run retains a denser and more persistent cold/molecular component than the late \high cloudy state, with high-density tails still extending to $n\gtrsim10^3\,{\rm cm^{-3}}$.

In the inner macro-scale ($1<r<10$~kpc, Figure~\ref{fig:pdf_phases_macro}), the phase PDFs trace primarily the radial growth and persistence of the multiphase medium rather than the strong density evolution seen in the nucleus.  At early times ($t/t_{\rm rain}=1$), both runs are almost entirely hot, with narrow hard/soft X-ray distributions and no significant cool component.  By $t/t_{\rm rain}\sim3$, only the \high run has already developed clear warm, cold, and molecular PDFs in this shell, indicating that the precipitation storm has expanded well beyond the central kpc. At later times ($t/t_{\rm rain}=5$--7$)$, both runs show an increasingly developed cool component, but the \high case remains strongly multiphase and more extended in density space, consistent with a more mature \cloudy phase reaching several kpc. In the \low run, the same radial range remains much more weakly multiphase at this stage, with only a limited warm/cold component and essentially no substantial molecular phase until the last time frame. Thus, in this radial range, the \cloudy regime is manifested mainly through the radial persistence of condensed gas, rather than through a marked late-time decrease in the characteristic densities of the cool phases.


\subsection{Micro to Macro evolution of gas masses}\label{subsec:masses}
\begin{figure}[!ht]
    \centering
    \includegraphics[width=0.94\linewidth]{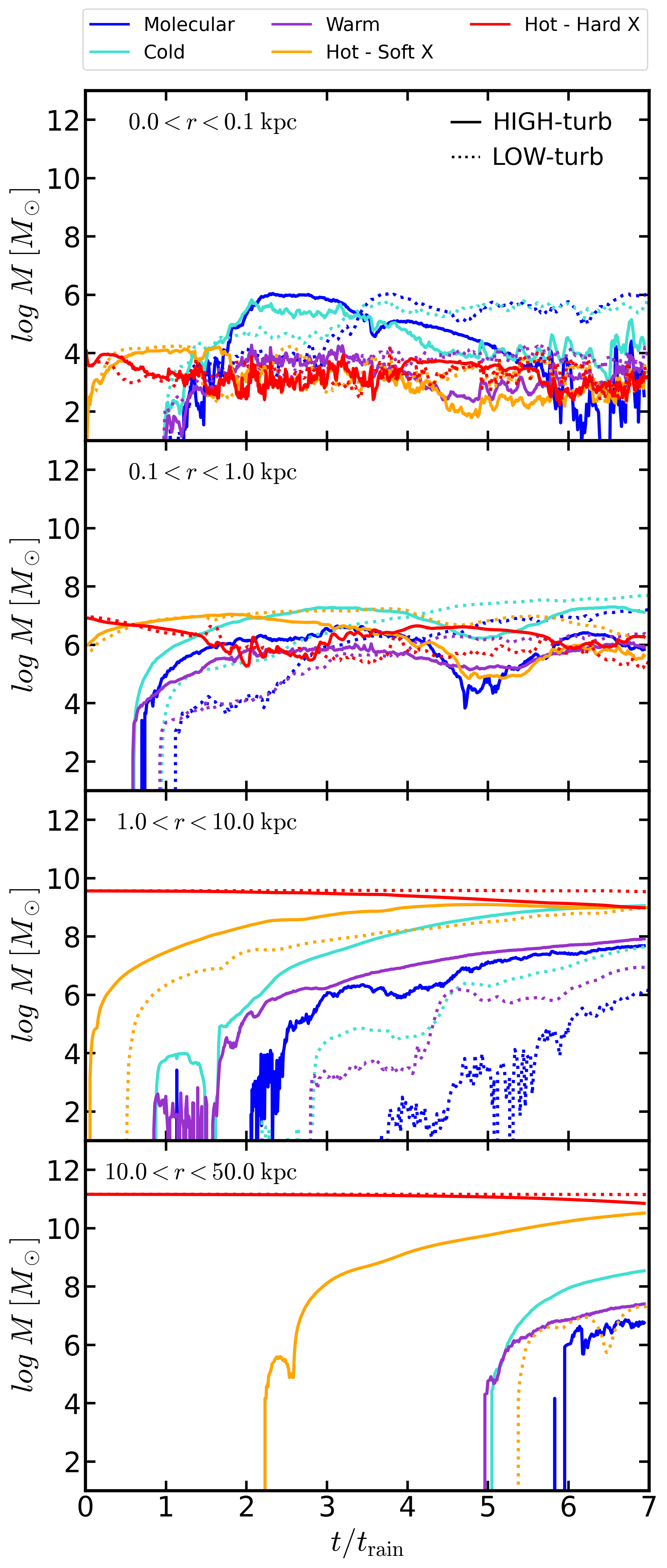}
    \caption{Phase masses as a function of time in the \low and \high case, as indicated in the legend, in different radial shells, from micro (top panel) to outer macro (bottom panel) according to the definitions in the text. The time is normalized to the corresponding \train in order to emphasize the weather cycles, starting from the activation of radiative cooling and jet feedback. For reference, \train $\simeq 9$ Myr in the \low run and $\simeq 16$ Myr in the \high run.}
    \label{fig:masses}
\end{figure}

Figure~\ref{fig:masses} shows the shell-integrated gas mass in each thermal phase as a function of time for the two jet-on reference runs, comparing the \high (solid lines) and the \low case (dotted lines). The figure provides a complementary view to the maps, profiles, and PDFs by directly tracking how the multiphase mass budget is assembled and redistributed across radius.  In all shells the hot phases dominate the total mass budget at early times, while the appearance of warm, cold, and molecular gas marks the onset of nonlinear cooling and condensation. The two turbulence regimes differ primarily in the timing of this transition and in how far from the center the condensed material is able to accumulate.

In the innermost shell ($0<r<0.1$~kpc), both runs develop a dense central reservoir shortly after $t/t_{\rm rain}\sim1$, but their subsequent evolution is markedly different. In the \high run, the molecular and cold masses rise rapidly and peak early with $M_{\rm mol/cold}\lesssim10^{6}\ M_{\odot}$, indicating prompt condensation and delivery of dense gas to the sink region. This early build-up is followed by a gradual decline and strong variability at later times, down to $M_{\rm mol/cold}\lesssim10^{2-4}\ M_{\odot}$ at $t/t_{\rm rain}\gtrsim5$, suggesting that the central reservoir is efficiently consumed, disrupted, or redistributed once jet activity becomes strong. In the \low run, by contrast, the central molecular/cold reservoir builds up more gradually ($M_{\rm mol/cold}\sim10^{4}\ M_{\odot}$ at $t/t_{\rm rain}\gtrsim1$) but remains more persistent, retaining a substantial mass to the latest times up to $M_{\rm mol/cold}\sim10^{6}\ M_{\odot}$. At fixed normalized time, the \high run therefore shows a more abrupt build-up of the inner cold reservoir, followed by faster depletion. In absolute time, however, this condensation still begins later than in the \low case since \train is a factor of $\sim2$ longer. We note that, in both runs, the warm gas mass closely resembles the cold gas mass trend, with a difference of about 2 orders of magnitude smaller, indicating a correlation due to the condensation cascade that bridges the two phases. The hot phases evolve more steadily at this scale, with $M_{\rm softX}$ and $M_{\rm hardX}$ remaining of order $10^{3-4}\,M_{\odot}$ in both the \high and \low runs.
By contrast, the molecular and cold components are much more variable, reaching $M_{\rm mol},M_{\rm cold}\lesssim10^{6}\,M_{\odot}$ during the main condensation episodes.

The same behavior extends to the $0.1<r<1$~kpc shell. Both runs populate this region with warm, cold, and molecular gas once precipitation begins, but the high-turbulence case again responds first, with a more rapid early growth of the dense phases up to $M_{\rm mol/cold}\gtrsim10^{6}\ M_{\odot}$ at $t/t_{\rm rain}\gtrsim1$.  At later times ($t/t_{\rm rain}\sim3$), however, the \low run catches up and becomes comparable to the \high case, or even exhibits larger amounts of mass for the cold and molecular components ($M_{\rm mol/cold}\lesssim10^{8}\ M_{\odot}$). Similar to the micro-scale, the warm gas mass follows the cold gas mass, up to $M_{\rm mol/cold}\sim10^{6}\ M_{\odot}$. Overall, this indicates that lower turbulence does not suppress condensation altogether; rather, after rain onset, it slows the outward growth of the condensed phase and keeps a larger fraction of the condensed gas confined to the inner kpc.

The clearest radial contrast appears outside the central kpc.  In the $1<r<10$~kpc shell, the \high run develops substantial warm, cold, and eventually molecular gas much earlier than the \low run, mainly due to the different strength of driven turbulence in the perturbed atmosphere, with masses as large as $M_{\rm mol/warm}\lesssim10^{8}\ M_{\odot}$ and $M_{\rm hot}\gtrsim10^{9}\ M_{\odot}$ at $t/t_{\rm rain}$ approaching 7. In the \low case, the same shell remains hot-dominated for longer, and the cool phases emerge only later and at remarkably lower masses ($M_{\rm mol}\sim10^{6}\ M_{\odot}$, $M_{\rm cold}\gtrsim10^{8}\ M_{\odot}$ and $M_{\rm warm}\sim10^{7}\ M_{\odot}$). The difference is even more pronounced in the outermost shell ($10<r<50$~kpc$)$, where only the \high run shows a clear late-time ($t/t_{\rm rain}\gtrsim5$) build-up of soft-X, warm ($M_{\rm warm}\sim10^{7}\ M_{\odot}$), cold ($M_{\rm cold}\gtrsim10^{8}\ M_{\odot}$), and molecular ($M_{\rm mol}\gtrsim10^{6}\ M_{\odot}$) material, while the \low run remains almost entirely dominated by the hot phase.  The hard X-ray component stays approximately steady and dominant at large radii in both runs ($M_{\rm hardX}\sim10^{11}\ M_{\odot}$), indicating that most of the outer atmosphere retains its diffuse hot character even as a multiphase component develops in the more turbulent case.

In general, Figure~\ref{fig:masses} reinforces the picture already suggested by the morphology and phase-space diagnostics: the jet-on \high atmosphere undergoes more radially extended multiphase development, whereas the \low run forms a denser but more centrally confined cold reservoir. In other words, stronger turbulence appears to favor the spatial spreading of precipitation, hence a transition from  \stormy to \cloudy phase, while weaker turbulence favors its central retention, which we define as \rainy. In addition, we note that with respect to B26a (see their Figure 11), where no-jet turbulence runs show a clear distinction at the meso-scale (with differences in the cooler gas masses up 2 orders of magnitude), in this work the meso/inner macro-scale condensation is partially supported by the jet, slightly erasing the difference between the two reference cases. This is particularly evident in the \low\ run, where the warm, cold, and molecular gas masses at meso/inner-macro scales are larger than in the corresponding no-jet B26a case by up to $\sim2$--$3$ dex, suggesting that jet-triggered condensation contributes substantially to CCA outside the nucleus.

\subsection{Jet power \& SMBH accretion history}\label{subsec:bhar}
\begin{figure}[!ht]
    \centering
    \includegraphics[width=1\linewidth]{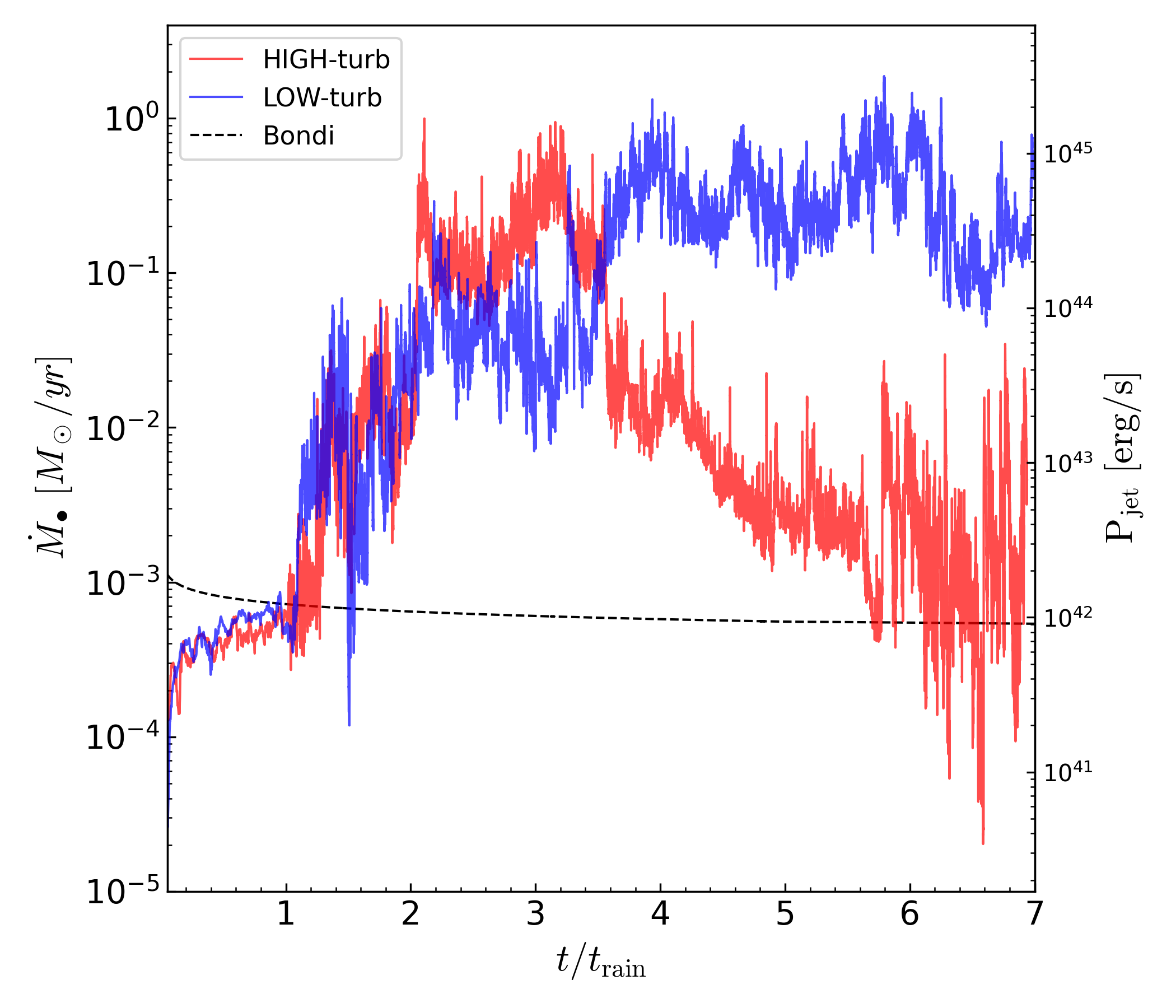}
    \caption{Black-hole accretion rate as a function of time in the \low and \high runs, as indicated in the legend. The time is normalized to the corresponding \train ($\sim 9$ and $\sim 16$ Myr for the \low and \high runs, respectively), starting from the activation of radiative cooling and jet feedback, in order to highlight the weather-cycle evolution. Despite the presence of the jet, \bhar reaches relatively large values, comparable to the no-jet B26a models shown by the dotted lines. The dashed black line indicates the idealized pure Bondi accretion rate.}
    \label{fig:bhar}
\end{figure}

In this Section we start by analyzing the accretion onto the central SMBH \bhar calculated as the total mass accreted by the sink particle in a simulation timestep, as already introduced. Figure~\ref{fig:bhar} shows the SMBH accretion rate, $\dot M_\bullet$, for the \low\ and \high\ runs as a function of time normalized to $t_{\rm rain}$.
The left-hand axis reports the accretion rate, $\dot{M}_{\bullet}$, while the right-hand axis gives the corresponding jet power for the adopted feedback prescription (see \S\ref{sec:setup}); the dashed curve marks the Bondi-like hot-mode reference. During the first rain time, $t/t_{\rm rain}\lesssim1$, the two reference runs show very similar trends, with slowly rising accretion rates toward the Bondi value ($\dot{M}_{\bullet}\sim10^{-3}\ M_{\odot}/{\rm yr}$). At $t/t_{\rm rain}\lesssim1$ we see that the \high case results in a slightly lower accretion rate with respect to the \low one (a few $10^{-4}$-$10^{-3}\ M_\odot/{\rm yr}$), mostly due to the higher level of turbulence injected at large scales which mixes up the gas, partially preventing further central accretion. This indicates that in the early stages the evolution of the system is not dramatically affected by the jet injection. In fact, turbulence directly injected by the jet itself has not yet sufficiently evolved to impact the gas thermodynamics in the close vicinity of the sink. 


Once multiphase gas reaches the inner region ($t/t_{\rm rain}\gtrsim1$), the accretion rate departs strongly from the Bondi-like hot-mode estimate and becomes highly variable, consistent with precipitation-driven feeding. Warm and cold clumps formed within the inner $\lesssim0.1-1$ kpc cool rapidly, lose support, and intermittently reach the sink, producing bursty episodes of enhanced SMBH fueling. In the present paper we focus on the gross morphology of this variability, while its detailed timing and kinematic diagnostics are deferred to the companion study, C26b.
Compared with the no-jet B26a runs, the key difference is not only the level of the accretion enhancement, but its weather-dependent duty cycle. In B26a, once CCA is established, both the \high\ and \low\ runs remain recurrently super-Bondi despite their different cold-gas morphologies, indicating that the presence of multiphase gas dominates the feeding rate. Here, the self-regulated jet makes the distinction sharper: the \high\ run produces a short, burst-dominated feeding episode followed by a lower-efficiency \cloudy\ stage, whereas the \low\ run maintains a longer-lived central reservoir and therefore a more persistent \rainy\ feeding cycle.

Although both runs experience this transition, the temporal character of the feeding cycle is very different.  The \high run enters an early and relatively impulsive accretion episode: $\dot{M}_{\rm BH}$ rises rapidly after $t/t_{\rm rain}\sim1$--2, reaches a short-lived maximum around $t/t_{\rm rain}\sim2$--3, and then declines sharply. After this burst, the system settles into a much lower accretion state, interrupted only by intermittent spikes. This behavior is consistent with the phase-mass evolution discussed above, in which the high-turbulence run builds a central dense reservoir quickly but does not retain it for long. In effect, the same turbulence that accelerates condensation also makes the inner flow more bursty and more likely to get disrupted via gas dragging. Given the overall trend of the \high run, combined with the density and temperature projections shown in the previous section, we can define two different weather conditions. The phase before $t/t_{\rm rain}\sim3.5$ is well idealized as \stormy, where \bhar shows more sustained yet short lived accretion episodes. After $t/t_{\rm rain}\sim3.5$, the \bhar enters a so-called \textit{cloudy} phase, where cold clouds and filaments keep forming at the meso and micro scales, but are not able to reach the sink region. 

At fixed normalized time, the \low run instead shows a somewhat later but much more sustained feeding history.  After the initial rise, the accretion rate remains elevated for several rain times, with a broad, long-lived phase of strong variability around a significantly super-Bondi level.  This prolonged active state is naturally explained by the more persistent molecular/cold reservoir in the central $\lesssim1$~kpc seen in Figure~\ref{fig:masses}. In this regime the black hole continues to be supplied by dense gas over an extended period, rather than through a single dominant burst.

Because the jet power is directly tied to the accretion rate in the adopted model, it follows the same qualitative behavior.  The high-turbulence run produces an earlier but shorter-lived high-power episode (in normalized time), whereas the low-turbulence run sustains jet activity for longer and with a larger duty cycle at late times. In both cases the inferred jet power exceeds the Bondi-level expectation by a wide margin once the flow becomes multiphase, reaching the range characteristic of strong mechanical feedback.  The accretion history therefore supports the same general interpretation as the thermodynamic diagnostics: in the jet-on configuration, ambient turbulence controls not only when condensation begins, but also whether the resulting black-hole fueling is impulsive and burst-dominated or prolonged and reservoir-fed.

\subsection{Phase thermodynamics}\label{subsec:phase}
\begin{figure*}[!ht]
    \centering
    \includegraphics[width=0.95\linewidth]{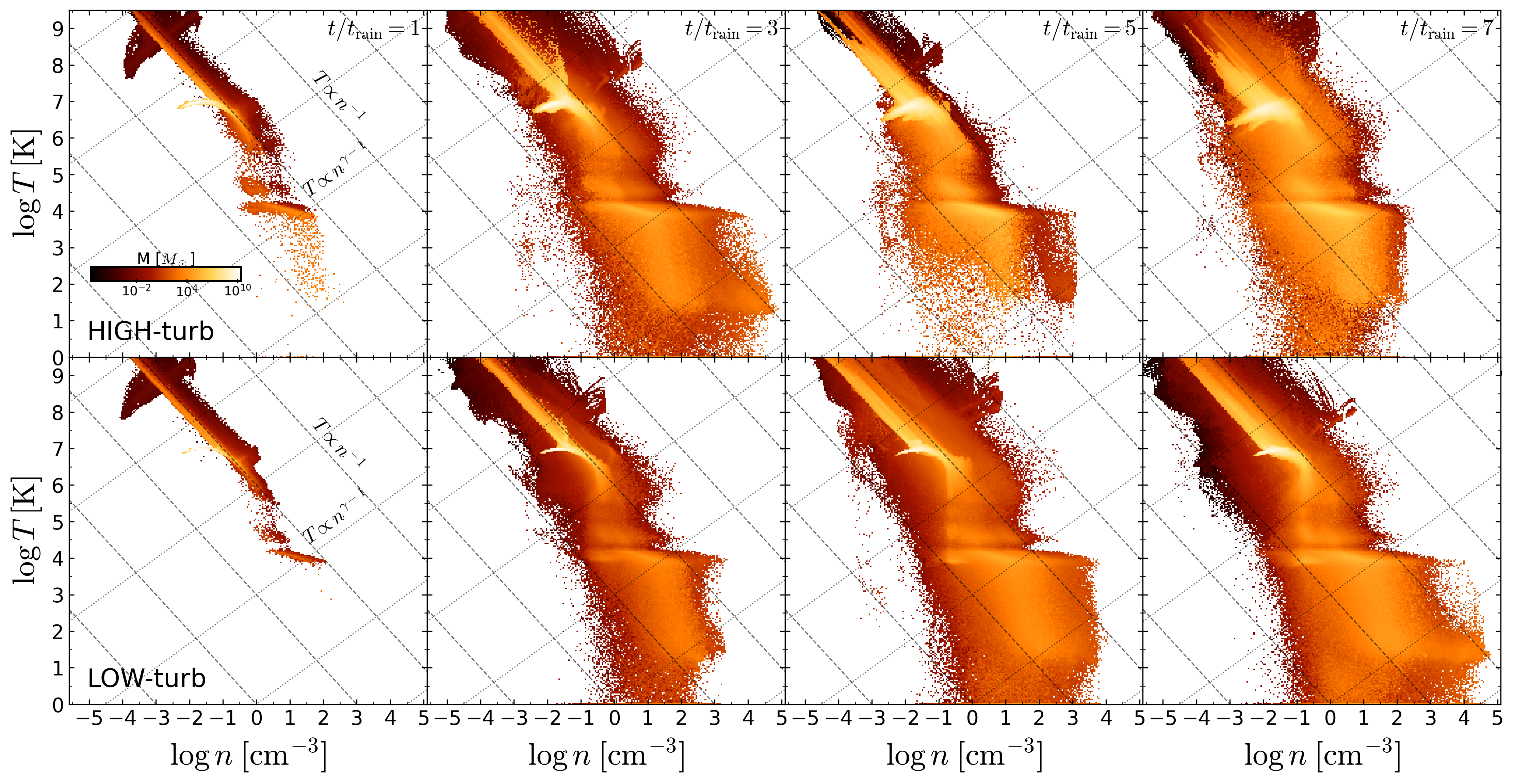}
    \caption{Density–temperature phase diagrams at four different epochs ($t/t_{rain}=1,3,5,7$) for the \high (first row) and \low (second row) reference runs. Axes show density and temperature; colors encode the gas mass per bin ($M_\odot$). At $t/t_{rain}=1$ the distributions are dominated by hot, rarefied gas displaced by the jet, with a prominent ridge following an approximately isobaric track ($T\propto n^{-1}$, denoted by the dashed lines). In the low-density/high-temperature region, a minor population of adiabatically expanding cells follows the dotted lines ($T\propto n^{\gamma-1}$). A minor but not negligible amount of gas has efficiently cooled, especially in the \high scenario where radiative losses populate a cool component near $T\sim10^{4}\ \mathrm{K}$, while the hot phase persists at lower densities. At $t/t_{rain}\gtrsim3$ the cooled, dense gas migrated to the center plausibly enhances accretion; the consequent jet power injection re-mixes and heats the surroundings, broadening the distribution toward higher $T$ and lower $n$ along with a few adiabatic expanding cells. Together, these stages illustrate the self-regulating CCA rain–feedback cycle: jet-driven uplift/mixing counteracts cooling on average, yet allows localized CCA condensation that feeds the central engine.}
    \label{fig:phase_plot_evo}
\end{figure*}

\begin{figure*}[!ht]
    \centering
    \includegraphics[width=0.95\linewidth]{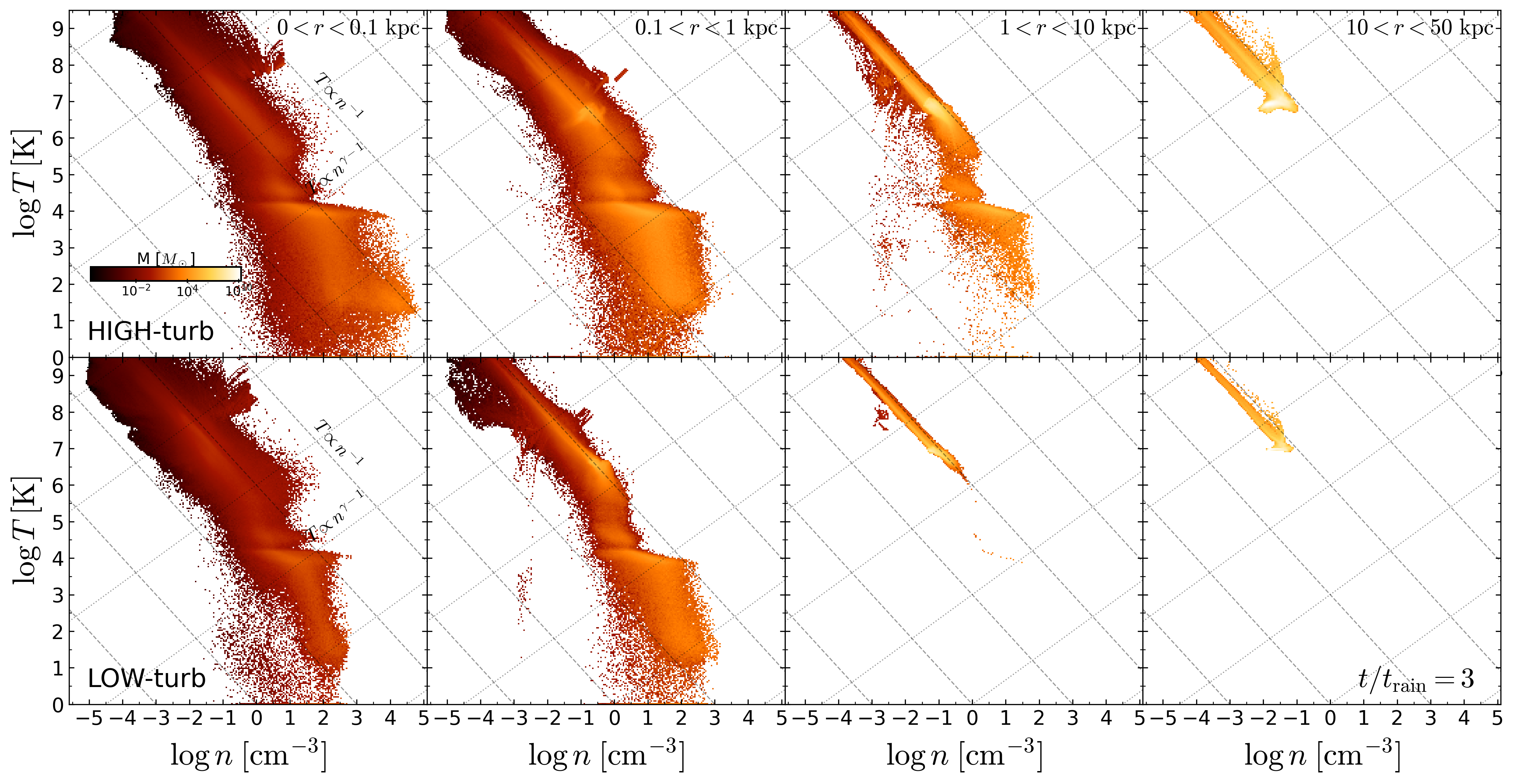}
    \includegraphics[width=0.95\linewidth]{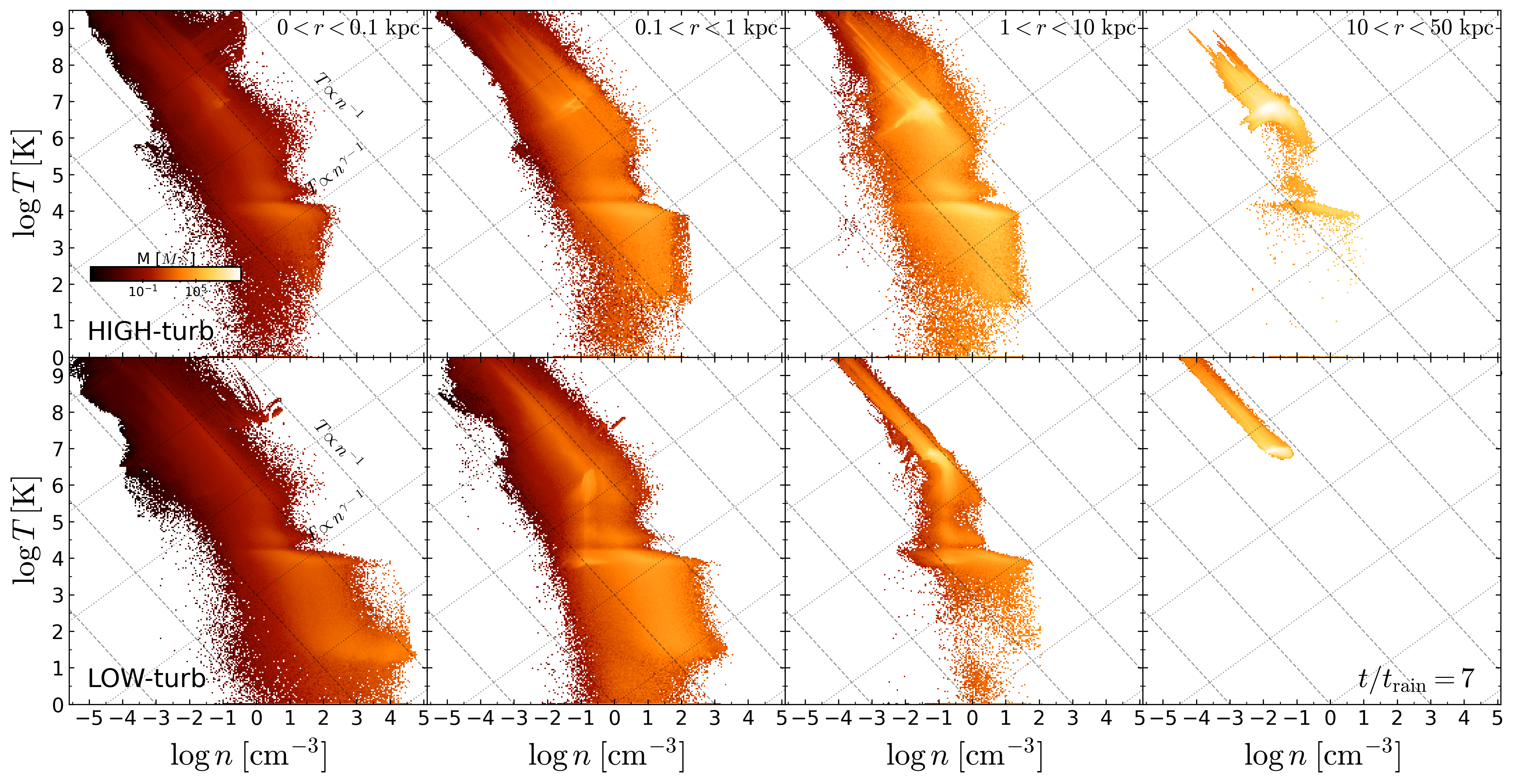}
    \caption{Density–temperature phase diagrams at four different radial bins at $t/t_{rain}=3$ (top figure) and $t/t_{rain}=7$ (bottom figure) from micro ($r<0.1$ kpc), meso ($0.1<r<1.0$ kpc), to inner ($1<r<10$ kpc) and outer ($10<r<50$ kpc) macro scales. Axes show density and temperature; colors encode the gas mass per bin ($M_\odot$).}
    \label{fig:phase_plot_radial}
\end{figure*}

Multiphase gas is ubiquitously observed in galaxy groups, as shown by several IGrM studies in the literature \citep{Sun2009, David2011, TemiAmblard2018, Eckert2021, OlivaresSalome2022, Temi2026}. In addition, recent exercises involving simulations have shown the presence of complex, multi-phase and multi-scale structures of gas in the hot halo of galaxy groups and clusters \citep[e.g. B26a,b, ][]{Fournier2025}. For instance, with turbulence stirring and radiative cooling only, B26a shows that, starting from a nearly hydrostatic equilibrium, the mono-phase gas develops significant multiphase states. In fact, these two ingredients alone trigger thermal instabilities able to further develop and turn into condensed structures following a non-linear evolution. This is especially favored by localized thermodynamic conditions, where cooling occurs on a timescale such that, compared to the typical free-fall time, $t_{\rm cool}/t_{\rm ff} \lesssim 10$ \citep{Field1965, Pizzolato2005, McCourt2012, Gaspari2017, Sharma2012}, effectively populating the cold-dense region of the phase plot. This is also supported by the condensation criterion introduced in \S\ref{sec:intro}, where gas is more prone to multiphase condensation and CCA formation when the cooling time is comparable, within a factor of 2, to the eddy turnover time $t_{\rm cool}\sim t_{\rm eddy}$ \citep[see B26a, b, ][]{Gaspari2018}.

In this section, we aim at describing how the injection of material in the jet geometry affects the overall distribution of the gas in terms of the temperature as a function of the number density, at different times and on different radial ranges, emphasizing the main differences with respect to the non-jet case.

\subsubsection{Phase structure and its time evolution}
\label{sec:phase_diagrams_time}
To connect the multi-scale morphology to the underlying thermodynamics, we show in Figure~\ref{fig:phase_plot_evo} the phase diagrams in the $(\log n,\log T)$ plane for the two reference runs (\high: top row; \low: bottom row) at four evolutionary stages ($t/t_{\rm rain}=1,3,5,7$ from left to right), color coded by the gas mass associated with each bin. At early times ($t/t_{\rm rain}=1$) both runs are dominated by the hot, volume-filling component, which occupies the low-density/high-temperature region and includes a very hot extension associated with jet-processed material.  As cooling proceeds, the distribution broadens along an approximately isobaric direction (diagonal guides with $T\propto n^{-1}$), indicating that a significant fraction of the gas cools while remaining close to pressure equilibrium with its surroundings. At $t/t_{\rm rain}\gtrsim3$, a substantial multiphase component appears, spanning the warm ($T\sim10^{4-5}$~K), cold ($T\lesssim10^{4}$~K), and molecular ($T\sim10^{1-2}$~K) regimes at progressively higher densities, and by $t/t_{\rm rain}=3$--4 the cold/molecular corner is strongly populated, indicating sustained condensation and fragmentation.

The subsequent evolution, however, differs between the two turbulence regimes. In the \high run, by $t/t_{\rm rain}\sim3$ the phase diagram becomes very broadly populated, with a continuous bridge connecting the hot atmosphere to the warm, cold, and molecular regimes over several orders of magnitude in density.  The intermediate-temperature phase is especially prominent, and the condensed component spans a wide range of densities rather than occupying a single, narrower branch. We interpret this trend as the thermodynamic signature of a genuinely \stormy precipitation phase: cooling, mixing, entrainment, and condensation proceed simultaneously, producing a highly disordered multiphase medium in which gas parcels sample a wide variety of thermal histories.  In the phase space, this corresponds to a broad and mildly separated distribution of hot, warm, and cold material.

At later times ($t/t_{\rm rain}=5$--7), the \high run remains strongly multiphase, but the distribution changes character. The low-temperature/high-density component becomes more clearly defined, while the phase-space bridge to the hot gas, although still present, appears less spread out than during the earlier \stormy stage. In addition, 
the cold and molecular phases occupy a narrower high-density range than during the earlier stormy stage, with a reduced extension toward the most extreme overdensities by $t/t_{\rm rain}=7$.
In fact, the condensed material increasingly resides in a better identified cold/dense wing of phase space, consistent with a transition toward a more \cloudy state: the atmosphere is still multiphase, but the condensed component is less reminiscent of a volume-filling precipitation storm and more of a population of dense clouds embedded in a hot, turbulent background.

The \low run evolves more gradually and with less phase-space broadening. Although it also develops a multiphase component by $t/t_{\rm rain}\gtrsim3$, the hot region remains narrower and the bridge toward the cold/molecular regime is less extended than in the \high case. The intermediate-temperature mixing layer is correspondingly less prominent, and the condensed material occupies a more coherent portion of the $(n,T)$ plane.  As the simulation evolves, it eventually produces a substantial cold component at late times ($t/t_{\rm rain}=5$--7), reaching densities comparable to the \stormy stage of the \high run at $t/t_{\rm rain}=3$. Given the smoother and constantly sustained evolution of the \low scenario, we tend to define the corresponding weather phase as \rainy. 

The two turbulence regimes differ primarily in \emph{how coherently} and \emph{fast} phase space is filled and how prominent the intermediate-temperature mixing layer becomes.  In the \high run, the phase distribution shows a wider variety of multiphase gas: at fixed temperature, it depicts asymmetric, wider, and skewed distributions from the hot to the cold/molecular components, consistent with stronger stirring, enhanced entrainment at interfaces, and a greater diversity of thermal histories.  The cold/molecular component results more dispersed in density and temperature at early times, while getting narrower in the \cloudy weather, which is the phase-space counterpart of the filamentary, fragmented morphology seen in the density projections shown in Figures~\ref{fig:mosaic_high} and \ref{fig:mosaic_low}.  In contrast, the \low run remains closer to a narrow hot locus at early times and develops a comparatively more concentrated condensed component at late times, with less extensive filling of the intermediate-temperature bridge; i.e. condensation is present but proceeds through a more coherent/less mixed pathway in phase space, resulting in progressively larger density values (from $n\lesssim 10^3 \ \rm{cm}^{-3}$ at $t/t_{\rm rain}\sim 3$ to $n\gtrsim 10^4 \ \rm{cm}^{-3}$ at $t/t_{\rm rain}\sim 5$--7). For $t/t_{\rm rain}\gtrsim 7$, the above described situation seems to have reverted, suggesting that the \high simulation has come to the end of a first accretion cycle, therefore returning to a narrower distribution, while the later stages of the \low run resemble the earlier diagrams of the \high run. Only the high-turbulence case shows a clear \stormy-to-cloudy transition: an early, strongly mixed precipitation episode followed by a later stage in which dense clouds make up a more distinct thermodynamic component.

\subsubsection{Radial stratification of phase space}
\label{subsec:phase_diagrams_radial}
We now analyze the radial-shell phase diagrams, which provide a more explicit view of how gas transitions across different atmospheres. Figure~\ref{fig:phase_plot_radial} complements the time sequence by showing the same $(\log n,\log T)$ mass distribution at two separate outputs, $t/t_{\rm rain}=3$ and 7, now separated into radial shells, i.e., micro- ($0<r<0.1$~kpc), meso- ($0.1<r<1$~kpc), inner macro- ($1<r<10$~kpc), and outer macro-scale ($10<r<50$~kpc) from left to right. The \high and \low runs are displayed in the top and bottom rows, respectively. Overall, the inner region ($r<0.1$~kpc) is strongly multiphase in both cases: hot gas coexists with a dense, cold/molecular component spanning several decades in density, and the condensed material preferentially lies along near-isobaric directions, consistent with rapid cooling of overdense structures embedded in a higher-pressure environment near the sink. Moving outward, the phase distribution becomes progressively dominated by the hot component in all scenarios, while the cold/molecular corner weakens, highlighting the spatial confinement of persistent rain of cold dense gas.

At $t/t_{\rm rain}=3$ (top figure), the \high run is in its \stormy phase. The inner shells ($r<0.1$~kpc and $0.1<r<1$~kpc) are already strongly multiphase, with hot gas coexisting with a broad warm/cold/molecular component spanning several decades in density.  We note that the $1<r<10$~kpc shell also shows a substantial multiphase signature: the hot branch is accompanied by an extended low-temperature, higher-density distribution, demonstrating that the precipitation storm is not confined to the nucleus but already extends out to the inner halo. By contrast, the \low run at the same epoch remains more centrally concentrated. While the inner kpc is multiphase, the $1<r<10$~kpc shell is much closer to a hot-dominated atmosphere with only a few cells in the cool component, and the outermost shell remains almost entirely hot. This comparison shows that high turbulence does not simply increase the amount of condensation, but also makes the \stormy phase radially more extended, confirming the trend observed in the previous sections.

By $t/t_{\rm rain}=7$, the \high run has evolved into a more \cloudy configuration. The inner two shells remain strongly multiphase, but the condensed component is now more clearly identifiable as a less dense, low-temperature branch rather than in the earlier stage, reaching densities of at most $n\lesssim 10^{2.5} \ \rm{cm}^{-3}$. The $1<r<10$~kpc shell still contains substantial cold/molecular material, indicating that the \cloudy state exhibits an even more radially extended and developed  condensation than at $t/t_{\rm rain}=3$: dense clouds and hot ambient gas are more widely distributed in phase space. Interestingly enough, in the outermost shell ($10<r<50$~kpc), a weak warm/cold tail becomes visible, consistent with uplifted or mixed material reaching large radii, even though the shell is still overwhelmingly dominated by diffuse hot gas. This may indicate that jet-driven turbulence has started to percolate to these scales, plausibly triggering condensation episodes.

The \low run also evolves between $t/t_{\rm rain}=3$ and 7, but along a different path. By the later epoch, multiphase gas has spread beyond the central kpc and the $1<r<10$~kpc shell now shows a clear cool component, indicating a delayed outward growth of the condensed phase with respect to the \high case. In this radial shell, the phase-space distribution remains narrower than in the \high run, and the outermost shell is still purely hot. In this sense, the \low atmosphere never develops a \stormy phase as broad as the \high case; instead, it moves more gradually from a centrally concentrated hot atmosphere toward a more modest and still relatively confined multiphase state, where the very dense gas in the innermost region provides fuel for more sustained accretion onto the center (see \S\ref{subsec:bhar}).

Together, the radial phase diagrams make the \stormy-to-cloudy transition in the \high run particularly clear.  At $t/t_{\rm rain}=3$, the atmosphere hosts a highly mixed radially extended precipitation storm, with multiphase gas already present out to several kpc.  By $t/t_{\rm rain}=7$, the same run is better described as \cloudy: dense condensed structures survive throughout the inner halo, but as a more distinct cold component, which shows less dense structures embedded within a still-dominant hot background.  The \low run follows the same general thermodynamic sequence, but with delayed outward growth, reduced radial extent, and a much weaker separation between an early \stormy phase and a later cloud-dominated one.

Overall, the difference between the \high and \low scenarios is most evident beyond the central kpc. In fact, in the \high run, the $1<r<10$~kpc shell still shows a clear multiphase signature, with a substantial amount of mass extending to warm/cold temperatures and moderate-to-high densities, implying that condensation and precipitation of cooled material is not limited to the micro/meso-scales but reaches several kpc. The big picture aligns with what is shown in B26a, where the general trend depicts a situation where the \high case develops larger and more variable condensed structures compared to the \low one, both in terms of spatial and thermodynamic ranges during the post-onset initial growth. 

An important difference from B26a is the value of \train\ in the \high\ case. While the \low\ run reaches precipitation on a similar absolute timescale in the jet-on and no-jet setups, the \high\ jet-on run develops rain roughly a factor of two earlier than the corresponding no-jet case in B26a. This suggests that jet-driven shear, compression, and turbulent mixing seed nonlinear perturbations that accelerate the post-onset development of multiphase rain, even though the \high\ run still condenses later than the \low\ run in absolute time.


\section{Discussion}\label{sec:disc}
The main goal of this work is to assess how CCA proceeds when radiative cooling and driven turbulence are supplemented by an actively responding, mass-loaded, kinetic jet in terms of morphological and thermodynamic properties. In B26a,b, the combination of turbulence and cooling alone already establishes the conditions under which a multiphase atmosphere can develop and condense. The present study extends that framework presented in B26a by adding mechanical feedback and by analyzing the resulting evolution through a deliberately multiscale and multiphase perspective. This combination is instrumental as the jet does not simply add heat to the background medium: it modifies the thermodynamic pathway of the gas, reshapes the spatial distribution of condensation sites, and ultimately changes how the condensed component couples back to black-hole feeding.

A first key outcome is that the jet does not merely advect pre-existing cold material, but reshapes when and where condensation occurs. Relative to the no-jet controls of B26a,b, the jet modifies the two turbulence regimes in complementary ways. In the \high case, the pre-existing density and entropy fluctuations provide abundant seeds for jet-driven compression, shear, and mixing, allowing multiphase gas to develop over a broader radial range and reducing the delay of precipitation relative to the corresponding no-jet run. Indeed, while in B26a,b the first cold clouds in the \high regime appear later than in the \low case by a factor of $\sim4$, here this contrast is reduced to $t_{\rm rain,high}\sim2\,t_{\rm rain,low}$. In the \low case, instead, the smoother background provides fewer nonlinear seeds, so the jet mainly introduces an additional condensation channel along the jet--atmosphere interface and organizes the cooled gas into fewer, more coherent structures that remain preferentially concentrated toward the center. Thus, the jet does not erase the B26a,b weather dichotomy, but amplifies it into distinct jet-regulated realizations: a more extended and mixed \stormy pathway in \high, and a more centrally retained \rainy pathway in \low.

The two turbulence regimes form distinct morphological pathways for jet-regulated precipitation. In the \high regime, the jet couples to a strongly perturbed medium, enhancing density contrast and producing an extended, fragmented filament network across scales. In the \low case, the jet acts on a smoother background, producing a comparatively well-organized transition towards a compact central cold structure. These trends highlight the multiscale nature of jet feedback: kpc-scale energy and momentum injection can imprint itself on pc-scale cloud formation, but the resulting morphology depends sensitively on the turbulent state that sets the initial spectrum of fluctuations and the efficiency with which jet-driven disturbances trigger local runaway cooling.

\subsection{CCA weather framework}
In this work, we use the weather labels in an operational sense. \textit{Stormy} denotes a phase with radially extended precipitation, broad phase-space occupation, strong mixing layers, and bursty central delivery. \textit{Rainy} indicates a more coherent and centrally confined condensation cycle, in which most of the cold phase remains within the inner kpc, a persistent central reservoir is maintained, and SMBH fueling stays elevated over a longer interval. \textit{Cloudy} denotes a later jet-processed configuration in which a substantial cold component still survives throughout the inner halo beyond the meso-scale, but is organized as a more distinct population of clouds and filaments embedded in a disturbed hot background. Finally, \textit{sunny} refers to a hot-dominated atmosphere in which cooling is temporarily subdominant to turbulent stirring and jet-driven clearing, where no cold clumps or filaments persist within a few kpc from the center and SMBH feeding stays near the Bondi-like hot-mode level.

The \high run appears to pass through an early \stormy phase in which precipitation is radially extended, highly mixed, and dynamically chaotic. In this regime, cooling, entrainment, uplift, and turbulent stirring all act together, so that the condensed gas is distributed across a broad density range and over a large radial extent. At later times, however, the same run evolves toward a different state: the condensed component remains substantial and radially extended, but it becomes more clearly identifiable as a population of less dense and sparse cold structures embedded in the hot background that struggle to lose angular momentum and to precipitate onto the central engine. We refer to this later stage as the \cloudy phase. 
Phase diagrams, accretion rates, PDFs, and shell-integrated masses all support this interpretation, showing that by late times the cold component in the \high run is more distinct in phase space and remains present out to larger radii, even though its presence at the micro scale results less dominant than during the \stormy stage. We note that additional stirring introduced by precessing jets may alter this evolution toward a completely \textit{sunny} phase, as observed in P26a,b, further limiting the central accretion.

The above interpretation may be useful for connecting CCA to the broader AGN feedback benchmark. In many precipitation-regulated and CCA-based models, cold gas formation is treated as the immediate response of the atmosphere to cooling once local thermodynamic conditions permit non-linear condensation. Our results suggest that, when a jet is present, this response may itself have internally distinct weather states. A \stormy state is characterized by vigorous multiphase rain, broad mixing layers, and bursty central delivery of condensed gas. Instead, a \cloudy state is characterized by a more distinct cloud population, continued but less volume-filling precipitation, and a more structured relation between dense gas, the central reservoir, and black-hole feeding. Also, the \rainy regime would  correspond to a more compact and relatively less extended cold gas structure, with high central activity. This distinction does not replace the standard CCA picture; rather, it refines it by highlighting that the same precipitation cycle may look qualitatively different depending on how the jet interacts with the ambient turbulent field.  In fact, when accounting for a multi-scale and multiphase approach, we appreciate that large-scale jet forcing is translated into local condensation through a cascade mediated by turbulence and cooling, showing how the CCA cycle must be understood across scales and phases at the same time.

The BH accretion histories further reinforce this diagnosis. Once condensation begins, both runs depart strongly from the Bondi-like hot-mode expectation, confirming that the feeding cycle is governed by multiphase precipitation rather than by smooth accretion from the hot background. However, the temporal character of the accretion differs sharply between the two weather regimes. The \high run exhibits more impulsive feeding, consistent with the chaotic formation and delivery of dense gas during the \stormy stage.  However, this feeding episode is comparatively short-lived, as the central reservoir is efficiently disrupted and redistributed via strong turbulence. On the other hand, the \low run develops a more persistent central cold reservoir, sustaining elevated accretion over a longer \train span. In this sense, the specific weather affects not only where precipitation occurs, but also whether black-hole fueling is bursty or prolonged, and therefore whether the jet duty cycle is dominated by short powerful episodes or by more sustained activity.

In addition, the phase-separated PDF shapes themselves shown in \S\ref{subsec:phase_diagrams_radial} carry useful physical information. Narrow, sharply peaked distributions indicate phases occupying a relatively well-defined thermodynamic state, as is often the case for the hot X-ray gas in the less disturbed atmosphere. By contrast, broad and asymmetric PDFs reflect a wider range of local conditions and thermal histories, consistent with turbulent stirring, jet-driven compression/rarefaction, and mixing across phase interfaces. In particular, the extended high-density tails of the cold and molecular phases during the peak precipitation epoch trace nonlinear condensation into compact overdensities, while their later broadening and mild shift toward lower densities suggest that these structures are no longer dominated by monotonic density growth, but instead by a balance of cloud formation, disruption, mixing, and accretion. The stronger overlap between adjacent phases in the \high run further supports the picture of a stormier, more dynamically coupled multiphase medium, whereas the narrower and more distinct phase PDFs in the \low case are consistent with a calmer and more coherent condensation cycle.

These findings resonate with the broader AGN feedback picture in which mechanical feedback both regulates and is regulated by multiphase condensation.  Observationally, radio-mode AGN systems often show hot X-ray cavities coexisting with warm filaments, molecular clouds, and evidence for uplifted or chaotically moving cold gas. The present simulation machinery provides a framework for interpreting such systems: the coexistence of hot cavities and cold precipitation suggests a self-regulated weather cycle in which the jet simultaneously disrupts and promotes condensation. From this perspective, the turbulence amplitude is a control parameter that determines how effectively jet feedback is converted into precipitation, entrainment, and cloud formation.

Finally, the present analysis also implies some caveats and directions for future work. We stress that the \cloudy phase introduced here should be regarded as a tentative phenomenological definition rather than a fully established dynamical category. A deeper investigation will require a wider exploration of the parameter space, including different jet powers, driving amplitudes, halo conditions, and possibly additional physics such as magnetic fields, conduction, or AGN winds.  Similarly, while the present simulations clearly show that the jet reshapes the CCA cycle, the detailed balance between uplift-triggered cooling, turbulent mixing, and direct compression may depend on the numerical setup and on the adopted subgrid feedback prescription. We plan to better characterize the implications described above in terms of kinematics and time variable quantities in the companion paper of this series, C26b, which will provide a complementary picture.

\subsection{A weather sequence for jet-regulated CCA}
\label{sec:weather_sequence}

\begin{figure*}
    \centering
    \includegraphics[width=0.75\textwidth]{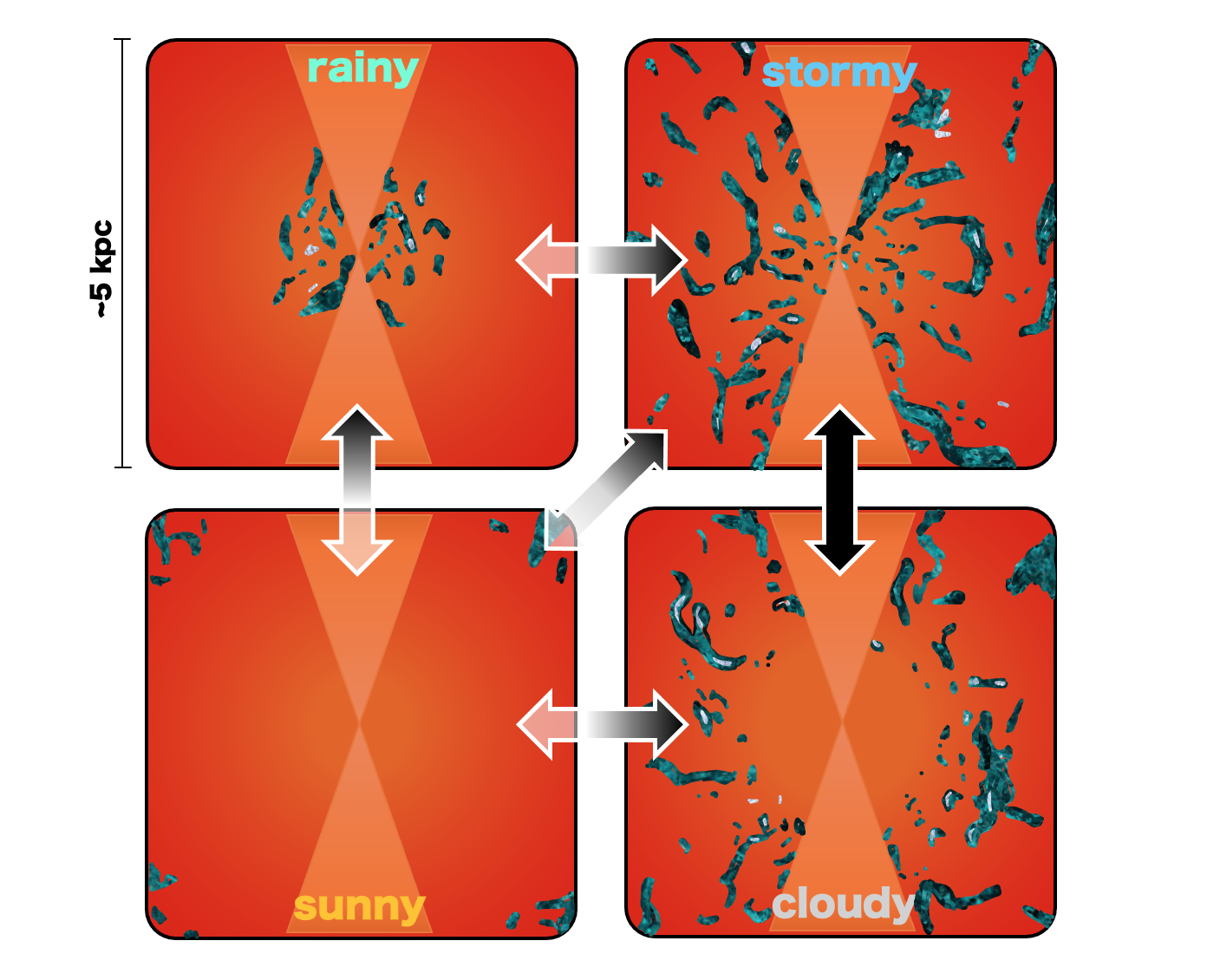}
    \caption{Schematic sketch of the proposed jet-regulated {\sc BlackHoleWeather} cycle. The four panels illustrate the phenomenological states \sunny, \stormy, \cloudy, and \rainy. \sunny: hot-dominated clearing stage, weak multiphase occupation, feeding near the Bondi-like baseline. \stormy: extended filamentary precipitation, broad hot--warm--cold phase-space bridge, burstier feeding. \cloudy: substantial cold gas survives as a more distinct cloud population embedded in a disturbed hot halo at the meso-scale, with reduced efficiency of central delivery. \rainy: more coherent and centrally concentrated precipitation, persistent inner reservoir, and longer-lived fueling. Black arrows indicate a qualitative feedback--feeding path, with their opacity schematically encoding the qualitatively favored transitions suggested by the present jet-regulated CCA runs: lower opacity marks less likely pathways.}
    \label{fig:weather_cycle_cartoon}
\end{figure*}

The results presented above suggest that, once micro-scale jet feedback is coupled to a turbulent, radiatively cooling atmosphere, the {\sc BlackHoleWeather} taxonomy should be extended beyond the \stormy/\rainy dichotomy discussed in pure cooling$+$turbulence setups (\citealt{Gaspari2013,Gaspari2015,Gaspari2017}; B26a,b). In particular, B26a,b provides the thermodynamic and morphological baseline for connecting turbulence strength to BH weather states: weak turbulence ($\mathcal{M}\sim0.15$) produces a compact, centrally concentrated \rainy configuration, whereas stronger turbulence ($\mathcal{M}\sim0.4$) delays condensation and promotes an extended, filament-rich \stormy state across meso-to-macro radii.  In the jet-on case, the atmospheric state is regulated not only by the onset of condensation, but also by how efficiently mechanical feedback redistributes low entropy gas, broadens the multiphase mixing layer, and modulates the delivery of condensed structures to the central engine. The two fiducial runs studied here therefore bracket different realizations of a common jet-regulated CCA cycle, rather than distinct accretion modes.

Within this framework, in Figure~\ref{fig:weather_cycle_cartoon} we try to provide a possible interpretation of the various weather states and transitions in the context of our reference simulations. The arrows indicate a possible evolutionary path, where darker colors indicate the qualitatively favored transitions suggested by the present simulations. In the fixed-axis jet-on runs studied here, a long-lived \sunny\ phase is not maintained. The system can briefly approach this state during the pre-rain or clearing intervals, but the fixed jet geometry does not efficiently remove the multiphase component from the full inner halo. By contrast, the spin-coupled runs of P26a,b show that jet reorientation can distribute feedback over a broader solid angle and produce longer central \sunny\ intervals: in that case, the inner $\sim100$ pc can become hot dominated even while fragmented multiphase gas persists at meso- and inner-macro radii.

A \stormy\ state is best represented by the \high\ run around $t/t_{\rm rain}\sim2$--$3$, when precipitation extends from the nucleus to several kpc and the phase diagrams, PDFs, and shell-integrated masses all show maximal multiphase breadth. A \rainy\ state is naturally associated with the \low\ run, where condensation starts earlier in absolute time, the hot cocoon remains more regular, and the multiphase medium is retained more efficiently in the central region. Finally, a \cloudy\ state emerges most clearly during the late evolution of the \high\ run, where the inner PDFs shift away from their most extreme overdensities, the cold branch becomes more distinct in phase space, and the accretion history declines from its earlier \stormy\ peak into a lower and more intermittent state. In this picture, the weather labels do not denote different accretion mechanisms, but different thermodynamic, morphological, and fueling manifestations of the same jet-regulated CCA cycle.


A plausible evolutionary interpretation is therefore the following. In a \rainy-like configuration, efficient central condensation builds a dense inner reservoir and sustains BH fueling, allowing prolonged jet activity. The resulting feedback inflates a hot channel, stirs the surrounding atmosphere, and can temporarily trigger a larger condensation efficiency along the jet-atmosphere interface. As low-entropy gas is uplifted, compressed, and mixed on larger scales, precipitation continues to radially extend, producing a \stormy phase with filamentary condensation across all scales. The continued feedback and turbulent percolation then reshape this precipitation pattern: the inner reservoir is progressively depleted or disrupted, the hot-warm-cold bridge narrows in the nucleus, and the atmosphere enters a \cloudy state in which the surviving clouds remain abundant beyond the inner kpc, but central fueling becomes less efficient. 
As stirring decays and cloud complexes rain back inward, the condensation region can contract again, returning the system toward a \rainy\ configuration or triggering a renewed precipitation episode.

At the same time, in our fixed jet-axis configuration, the weather is naturally less prone to evolve toward a \sunny\ state. In fact, without reorientation of the jet, cold clouds and filaments precipitating from the meso- to micro-scale are prevented from completely vanishing from the inner macro-scale, disfavoring a transition from \rainy/\stormy to \sunny. In addition, the reduced central activity typical of the \cloudy state naturally leads to a smaller efficiency in entrainment, uplift and reheating of cold condensates, especially beyond the inner kpc, making the transition to \sunny less likely. Conversely, in spin-dependent runs (see P26a,b), the jet reorientation and precession can efficiently drive the system towards a \sunny weather even from a \stormy state. We stress that this sequence should be regarded as a physically motivated phenomenological framework rather than as a fully demonstrated closed loop. Our simulations adopt fixed turbulence driving and a fixed jet prescription, and therefore do not yet self-consistently follow all transitions within a single realization (see also P26a,b). Nevertheless, the results presented here already show that moderate changes in the ambient turbulent state, once coupled to jet feedback, are sufficient to shift the same hot halo between an extended \stormy precipitation regime, a centrally confined \rainy regime, and a later \cloudy state. The proposed weather sequence therefore provides a useful physical language for linking multiphase morphology, phase-space structure, and SMBH fueling within the broader {\sc BlackHoleWeather} program, while also motivating future simulations with time-dependent driving, variable jet geometry, magnetic fields, and BH spin coupling.

\subsection{Comparison with previous works}
\label{sec:comparison_simulations}

The present study lies at the intersection of two simulation strands that have often been explored separately. On the one hand, early self-regulated mechanical-feedback calculations showed that bipolar kinetic outflows can suppress catastrophic cooling while preserving the global cool-core structure in groups and clusters, with lower-mass halos requiring a gentler and more persistent coupling than massive cluster atmospheres \citep{GaspariBrighenti2011,GaspariMelioli2011,Gaspari2012}. On the other hand, pioneering CCA simulations \citep{Gaspari2013,Gaspari2015,Gaspari2017} established that, in a turbulent and radiatively cooling halo, nonlinear condensation drives stochastic cold inflow, broad multiphase structure, and BH accretion rates well above the Bondi prediction. Relative to these earlier works, this work combines both ingredients in a single hyper-zoom group-halo setup: a self-regulated micro-scale kinetic jet evolves together with radiative cooling and controlled ambient turbulence, using B26a,b as the no-jet reference. The main new result is that once a jet is present, the same CCA evolution can be displaced between \stormy, \rainy and late \cloudy\ realizations, depending on the turbulent state of the ambient medium.

Our results are broadly consistent with earlier AGN mechanical-feedback calculations. In particular, \citet{GaspariBrighenti2011,GaspariMelioli2011,Gaspari2012} showed that bipolar mechanical outflows can self-regulate cooling in groups and clusters while preserving the cool-core structure, with the coupling remaining intrinsically anisotropic and more delicate in lower-mass halos than in massive clusters. Jet-driven heating reshapes the thermodynamic state of the core through directional energy deposition, circulation, weak shocks, uplift, and mixing. Similarly, \citet{Yang2016,Yang2019} emphasize that jet heating is intrinsically anisotropic, with entropy gain concentrated inside the jet cones and the global core maintained through a combination of shocks, mixing, adiabatic processes, and circulation rather than through turbulent dissipation alone. Likewise, \citet{Prasad2015} and \citet{Li2014} find cold-gas/jet hysteresis cycles and stochastic feeding in cluster cores. In addition, \citet{WittorGaspari2020} investigate AGN-driven turbulent weather in self-regulated feedback simulations, showing recurrent strong and weak turbulence episodes over long-term evolution, while \citet{WittorGaspari2023} show that baroclinicity can become important at and below the meso scale during active jet phases, triggering fresh turbulence in stratified atmospheres. The presented work corroborates these results in three key respects: jets do not simply erase cooling, the thermodynamic response is highly anisotropic and multiphase, and SMBH fueling remains bursty and cold-gas regulated. The main difference is that here the ambient turbulent state itself is isolated as a physical control parameter. At similar evolutionary stage, the \high run develops an extended \stormy precipitation pattern followed by \cloudy reorganization, whereas the \low run remains in a more centrally confined \rainy regime.

A second point of comparison concerns the internal structure of the cold phase and its interaction with the jet. In the no-jet B26a setup, stronger stirring already produced a broader, more extended precipitation pattern, while weaker stirring yielded a more centralized and coherent \rainy configuration. The present paper shows that this distinction is not erased by feedback; rather, the jet amplifies it into a richer weather sequence. At fixed normalized time, the \high run develops a broader hot--warm-cold span, a larger radial extent of the cold phase, and a more porous cocoon, while the \low run retains a more regular hot channel and a longer-lived inner cold reservoir. This work extends the CCA weather framework of B26a,b from a pure cooling$+$turbulence regime to a genuinely jet-regulated one. At the same time, the late-time \cloudy state identified here has no direct analogue in the simpler no-jet setup, because it emerges from the combined action of precipitation, jet-driven redistribution, and partial depletion of the central reservoir.

Recent MHD calculations indicate that additional microphysics can further reshape the weather sequence introduced above. Using \athenapk, \citet{Fournier2024} showed in Perseus-like cluster simulations that magnetic fields suppress the formation of massive cold disks \citep[see also][]{Ehlert_2022}, favor magnetically supported filaments over clumpy structures, and can recurrently deflect jets through their interaction with dense cold material. In this context, the present calculations should be regarded as a controlled hydrodynamic baseline. At the same time, the jet injection setup is very different: their lighter, cluster-scale jet efficiently excavates a persistent channel, which can be problematic for maintaining strong coupling to the surrounding multiphase medium over long times. Conversely, our heavier, group-scale mass-loaded jet remains more strongly coupled to the ambient gas. The large-scale weather sequence and its dependence on ambient turbulence are therefore likely robust, but the detailed jet-filament coupling, the survival of the smallest cold structures, and the compactness of the inner reservoir may change once magnetic fields are included.

In addition, this work is complementary to simulations that push inward toward the Bondi and horizon scales rather than outward through the full halo weather cycle. In nested-mesh hydrodynamic calculations of M87-like elliptical systems, \citet{Guo2023} followed radiative cooling and phenomenological heating across several orders of magnitude in radius, finding that multiphase accretion can persist down to near-horizon scales. Likewise, the multizone GRMHD technique of \citet{Cho2024} was explicitly designed to bridge Bondi-scale and horizon-scale accretion over a very large dynamic range. While those calculations capture the inner flow with higher physical fidelity, they do not follow a precipitation-regulated group atmosphere with a realistic self-regulated jet-weather coupling across the full halo.

Within the BHW paper sequence, B26a therefore provides the controlled no-jet limit, where turbulence alone separates extended \stormy\ from compact \rainy\ CCA. The present work shows that a fixed-axis kinetic jet does not erase this dichotomy, but reshapes it by lowering the peak central cold densities, redistributing condensed gas along the jet--ambient interface, and introducing a late \cloudy\ state with inefficient central delivery. P26a then add the next layer of physics: spin-coupled jet reorientation changes the feedback geometry, allowing longer central \sunny\ intervals and making the long-term weather cycle sensitive not only to feedback power, but also to the directionality and coherence of the feeding--feedback coupling.

Overall, the broader simulation literature supports the main physical picture emerging here: self-regulated mechanical feedback and multiphase precipitation are robust outcomes of radiatively cooling hot halos, but their detailed manifestation depends sensitively on the injected turbulence, the jet model, and the included microphysics. Within this landscape, this paper suggests that the meso-to-micro feedback bridge in a group-scale atmosphere can support distinct weather states with measurable differences in morphology, phase-space structure, and SMBH fueling.

\section{Conclusions}
\label{sec:concl}

In this work, we investigate how AGN jet feedback reshapes the morphology and thermodynamics of chaotic cold accretion (CCA) in a turbulent, radiatively cooling group atmosphere. Building on the cooling$+$turbulence baseline established in B26a,b and \citet{GaspariBrighenti2011,Gaspari2013,Gaspari2017}, we presented the jet-on extension of the same multiscale {\sc BlackHoleWeather} framework, focusing on how feedback modifies the condensation cycle once the atmosphere is already able to produce multiphase rain. To this end, we analyzed two reference hydrodynamical simulations at sub-pc resolution ($\Delta x_{\rm min}\simeq 0.78$ pc), well below the Bondi radius, including radiative cooling, driven subsonic turbulence, and a self-regulated kinetically-dominated and mass-loaded jet. The two runs differ only in their ambient turbulence level, allowing us to isolate how strongly the turbulent state of the halo controls the subsequent feedback-condensation cycle.

Both simulations activate radiative cooling and jet feedback in an already developed turbulent atmosphere and then evolved through several precipitation episodes starting from a hot, stratified IGrM halo. The main advance of this paper is to show that, once a self-regulated jet is included, CCA is not suppressed into a featureless heating state; rather, it is reorganized into distinct weather realizations whose morphology, phase-space structure, and fueling properties depend sensitively on the ambient turbulent field. Our main conclusions are as follows.
\begin{enumerate}
    \item The AGN jet does not simply offset cooling through heating. Instead, it reshapes the CCA cycle through compression, shear, entrainment, and turbulent percolation. Its coupling to the atmosphere is intrinsically \textit{anisotropic}: feedback acts most directly along the jet channel and its interface, while condensation is redistributed and locally promoted rather than uniformly suppressed. Feedback is therefore not external to precipitation, but part of the same self-regulated baryon cycle that governs where and when nonlinear condensation develops.\\

    \item The ambient turbulent state is a key control parameter of jet-regulated CCA. In the \high run, condensation starts later in absolute time ($t_{\rm rain}\simeq 16$ Myr), but once established it becomes more radially extended, filamentary, fragmented, and strongly mixed. In the \low run, condensation starts earlier ($t_{\rm rain}\simeq 9$ Myr) and remains more coherent and centrally confined, with a larger fraction of the cooled material retained within the inner kpc. The no-jet weather dichotomy identified in B26a,b is therefore not erased by feedback, but amplified into distinct jet-regulated realizations.\\

    \item The \high case shows a clear transition from an early \stormy phase to a later \cloudy phase. The \stormy stage is marked by broad phase-space occupation, strong intermediate-temperature mixing, filamentary precipitation extending to several kpc, and burst-dominated fueling. At later times, the condensed component remains substantial and radially extended, but becomes more clearly identifiable as a distinct cold/dense cloud population embedded in the hot halo, with weaker central delivery and reduced dominance of the most extreme inner overdensities. By contrast, the \low case remains closer to a persistent \rainy cycle, with a more compact morphology, a more regular hot cocoon, a narrower phase-space bridge, and a longer-lived central cold reservoir.\\

    \item The multiscale diagnostics provide a consistent physical picture of jet-regulated CCA across the macro, meso, and micro scales. The maps reveal the weather-dependent fragmentation of the cold gas and the differing cocoon geometries; the radial profiles show the thermodynamic and kinematic separation between the extra-cone atmosphere and the intra-cone hot channel; the PDFs and phase diagrams quantify the broadening, mixing, and radial spread of the condensed component; and the shell-integrated phase masses show that stronger turbulence favors a wider spatial distribution of precipitation, whereas weaker turbulence favors central retention. These diagnostics indicate that the meso scale is an active processing layer where jet feedback reorganizes halo rain both geometrically and thermodynamically: persistent condensation is suppressed inside the hot, rarefied jet channel and survives preferentially in the surrounding atmosphere and along the jet-ambient interface.\\

    \item Black-hole fueling is tightly coupled to weather state. Once condensation begins, both runs depart strongly from the Bondi-like hot-mode expectation, confirming that accretion is CCA-driven rather than controlled by the smooth hot background. The \high run shows an abrupt, burst-dominated feeding episode during the \stormy stage, followed by a lower and more intermittent \cloudy regime. By contrast, the \low run maintains a longer-lived central cold reservoir and sustains enhanced accretion and jet activity over a broader time interval, consistent with a more persistent \rainy cycle. Turbulence therefore regulates not only where multiphase gas forms, but also whether feeding is impulsive or reservoir-fed.

\end{enumerate}

These results strengthen the broader {\sc BlackHoleWeather} interpretation of AGN feeding and feedback as a genuinely multiscale weather cycle. The jet and the multiphase atmosphere are not separate components, but two parts of the same self-regulated system: the jet stirs, compresses, uplifts, and redistributes the gas in an anisotropic way, thereby influencing where condensation occurs, while the condensed gas in turn fuels the SMBH and modulates the subsequent feedback power. In this framework, turbulence sets atmospheric conditions in different weather realizations, determining whether the system develops an extended \stormy precipitation pattern, a centrally retained \rainy configuration, or a later \cloudy state with inefficient central fueling.

Overall, this work moves CCA beyond the pure cooling$+$turbulence limit into the jet-regulated regime, where feedback and precipitation become key ingredients of the same multiscale baryon cycle. By showing how ambient turbulence selects distinct weather realizations and how these imprint on morphology, phase structure, and SMBH fueling, this work establishes the physical bridge between halo-scale thermodynamics, meso-scale multiphase processing, and micro-scale feeding. In this picture, it provides a core pillar of the {\sc BlackHoleWeather} program: a predictive framework in which AGN feedback is interpreted not as static heating, but as a time-dependent weather engine coupling the hot halo to the black hole across scales.\\

\begin{acknowledgements}
The {\sc BlackHoleWeather} authors acknowledge key funding support from the European Research Council (ERC) under the European Union's Horizon Europe research and innovation programme (Consolidator Grant {\sc BlackHoleWeather}, No.~101086804; PI: Gaspari). Views and opinions expressed are, however, those of the author(s) only and do not necessarily reflect those of the European Union or the European Research Council Executive Agency; neither the European Union nor the granting authority can be held responsible for them. 
We acknowledge ISCRA for awarding this project access to the LEONARDO supercomputer, owned by the EuroHPC Joint Undertaking, hosted by CINECA (Italy).
We acknowledge EuroHPC Joint Undertaking for awarding us access to Karolina at IT4Innovations (Czech Republic) and MeluXina at LuxProvide (Luxembourg).
The numerical work was in part supported by the NASA High-End Computing (HEC) Program through the NASA Advanced Supercomputing (NAS) Division at Ames Research Center. 
VO acknowledges support from the DICYT ESO-Chile Comite Mixto PS 1757, Fondecyt Regular 1251702, and CIRAS-AI Project, code FIUF137139-USACH.
FMM carried out part of the research activities described in this paper with contribution of the Next Generation EU funds within the National Recovery and Resilience Plan (PNRR), Mission 4 - Education and Research, Component 2 - From Research to Business (M4C2), Investment Line 3.1 - Strengthening and creation of Research Infrastructures, Project IR0000034 – “STILES - Strengthening the Italian Leadership in ELT and SKA.
PT acknowledges support from NASA NNH22ZDA001N Astrophysics Data and Analysis Program under award 24-ADAP24-0011.
We thank Amirnezam Amiri and Fabrizio Fiore for comments and discussions.
We thank Philipp Grete for support with the \athenapk\ code.
We thank the organizers and participants of the following conferences for the stimulating discussions that helped improve this work: `BlackHoleWeather I' (Sexten, ITA), `SMBH-2025' (G\"oteborg, SWE), and `Breaking the Limits 2026' (Cagliari, ITA).
\end{acknowledgements}

\bibliography{biblio/my}{}
\bibliographystyle{aa}

\appendix
\onecolumn
\section{PDF statistics}
Statistical properties of the Probability Density Functions (PDFs) presented in \S\ref{subsec:stats} are shown in Table~\ref{tab:pdf_moments_all}.  
\begin{table*}[!htbp]
\centering
\footnotesize
\setlength{\tabcolsep}{4pt}
\renewcommand{\arraystretch}{1.1}
\begin{tabular*}{\textwidth}{@{\extracolsep{\fill}}lllcccccccccccc@{}}
\hline
\textbf{Run} & $t/t_{\rm rain}$ & \textbf{Phase} & \multicolumn{3}{c}{$0<r<0.1$ kpc} & \multicolumn{3}{c}{$0.1<r<1$ kpc} & \multicolumn{3}{c}{$1<r<10$ kpc} \\
&  &  & $\mu_{\log n}$ & $\sigma_{\log n}$ & $\mathcal{S}_{\log n}$ & $\mu_{\log n}$ & $\sigma_{\log n}$ & $\mathcal{S}_{\log n}$ & $\mu_{\log n}$ & $\sigma_{\log n}$ & $\mathcal{S}_{\log n}$ \\
\hline
\noalign{\vskip 3pt}
 &  & Molecular & --- & --- & --- & 1.825 & 0.297 & -0.978 & --- & --- & --- \\
 &  & Cold & 1.289 & 0.247 & -0.359 & 1.227 & 0.421 & -0.623 & 1.433 & 0.384 & -0.659 \\
\high & 1 & Warm & 0.560 & 0.282 & -0.351 & -0.140 & 0.185 & 0.780 & 0.099 & 0.077 & -0.364 \\
 &  & Hot - Soft X & -0.608 & 0.084 & 1.832 & -0.675 & 0.104 & 0.771 & -0.836 & 0.063 & 0.694 \\
 &  & Hot - Hard X & -0.830 & 0.228 & -1.575 & -0.875 & 0.076 & -7.310 & -1.229 & 0.091 & 0.283 \\
\hline
\noalign{\vskip 3pt}
 &  & Molecular & 3.820 & 0.695 & -1.235 & 1.713 & 0.430 & -0.489 & 1.165 & 0.520 & -0.769 \\
 &  & Cold & 2.151 & 0.750 & 0.945 & 1.055 & 0.463 & -0.038 & 0.711 & 0.435 & -0.187 \\
\high & 3 & Warm & 0.676 & 0.363 & 0.012 & -0.075 & 0.332 & 0.049 & -0.364 & 0.263 & 0.007 \\
 &  & Hot - Soft X & -0.117 & 0.409 & 0.923 & -0.808 & 0.185 & 0.566 & -0.995 & 0.138 & 0.403 \\
 &  & Hot - Hard X & -0.740 & 0.566 & 0.779 & -0.996 & 0.345 & -2.031 & -1.220 & 0.165 & -3.588 \\
\hline
\noalign{\vskip 3pt}
 &  & Molecular & 2.718 & 0.241 & -1.153 & 1.298 & 0.521 & -1.138 & 0.756 & 0.553 & -1.302 \\
 &  & Cold & 2.093 & 0.780 & -0.586 & 0.734 & 0.427 & -0.275 & 0.353 & 0.419 & -0.324 \\
\high & 5 & Warm & 0.225 & 0.343 & 0.286 & -0.256 & 0.286 & -0.760 & -0.629 & 0.263 & -0.475 \\
 &  & Hot - Soft X & -0.939 & 0.282 & 1.844 & -1.161 & 0.307 & 0.364 & -1.209 & 0.246 & -0.511 \\
 &  & Hot - Hard X & -1.240 & 0.120 & -5.458 & -1.267 & 0.116 & -5.404 & -1.327 & 0.163 & -1.395 \\
\hline
\noalign{\vskip 3pt}
 &  & Molecular & 1.320 & 0.518 & -1.334 & 1.182 & 0.483 & -0.949 & 0.556 & 0.507 & -1.017 \\
 &  & Cold & 1.041 & 0.449 & -0.247 & 0.712 & 0.450 & -0.233 & 0.117 & 0.432 & -0.205 \\
\high & 7 & Warm & 0.167 & 0.445 & -0.124 & -0.153 & 0.448 & -0.255 & -0.767 & 0.335 & -0.477 \\
 &  & Hot - Soft X & -0.684 & 0.614 & 1.293 & -0.804 & 0.622 & 0.328 & -1.393 & 0.250 & -0.497 \\
 &  & Hot - Hard X & -1.320 & 0.388 & -0.479 & -1.269 & 0.286 & -0.133 & -1.460 & 0.184 & -1.673 \\
\hline
\noalign{\vskip 3pt}
 &  & Molecular & --- & --- & --- & --- & --- & --- & --- & --- & --- \\
 &  & Cold & 1.386 & 0.292 & -0.027 & 1.235 & 0.369 & 0.151 & --- & --- & --- \\
\low & 1 & Warm & 0.295 & 0.259 & 0.730 & 0.054 & 0.188 & 0.764 & --- & --- & --- \\
 &  & Hot - Soft X & -0.492 & 0.096 & 1.957 & -0.724 & 0.071 & 1.832 & -0.805 & 0.014 & 0.301 \\
 &  & Hot - Hard X & -0.777 & 0.233 & -1.719 & -0.876 & 0.146 & -3.323 & -1.220 & 0.089 & 0.956 \\
\hline
\noalign{\vskip 3pt}
 &  & Molecular & 2.255 & 0.398 & -0.163 & 1.964 & 0.425 & -0.704 & --- & --- & --- \\
 &  & Cold & 1.632 & 0.538 & 0.108 & 1.253 & 0.501 & -0.484 & 1.236 & 0.432 & 0.010 \\
\low & 3 & Warm & 0.318 & 0.429 & 0.468 & -0.238 & 0.275 & 1.480 & -0.039 & 0.097 & 0.547 \\
 &  & Hot - Soft X & -0.639 & 0.602 & 1.693 & -0.551 & 0.100 & 0.179 & -0.773 & 0.081 & 0.572 \\
 &  & Hot - Hard X & -1.521 & 0.450 & 0.495 & -1.087 & 0.347 & -1.169 & -1.183 & 0.123 & -2.131 \\
\hline
\noalign{\vskip 3pt}
 &  & Molecular & 2.700 & 0.490 & -0.264 & 1.795 & 0.457 & -0.628 & 0.711 & 0.320 & -1.131 \\
 &  & Cold & 2.170 & 0.596 & -0.018 & 1.034 & 0.595 & -0.539 & 0.617 & 0.516 & -0.684 \\
\low & 5 & Warm & 0.808 & 0.479 & -0.627 & -0.248 & 0.417 & 1.034 & -0.499 & 0.246 & 1.078 \\
 &  & Hot - Soft X & 0.214 & 0.626 & 0.258 & -0.541 & 0.176 & 0.508 & -0.806 & 0.119 & 1.283 \\
 &  & Hot - Hard X & -0.587 & 0.648 & 0.368 & -1.346 & 0.479 & -0.508 & -1.159 & 0.141 & -3.147 \\
\hline
\noalign{\vskip 3pt}
 &  & Molecular & 3.524 & 0.761 & -0.635 & 1.642 & 0.509 & -0.545 & 1.079 & 0.628 & -0.437 \\
 &  & Cold & 1.936 & 0.794 & 0.874 & 0.981 & 0.618 & -0.619 & 0.474 & 0.649 & -0.312 \\
\low & 7 & Warm & 0.307 & 0.432 & -0.109 & -0.518 & 0.418 & 0.983 & -0.646 & 0.228 & 1.631 \\
 &  & Hot - Soft X & -0.638 & 0.528 & 1.205 & -0.753 & 0.180 & 1.337 & -0.934 & 0.136 & -0.168 \\
 &  & Hot - Hard X & -0.835 & 1.008 & 0.318 & -1.594 & 0.516 & 0.060 & -1.168 & 0.116 & -1.270 \\
\hline
\end{tabular*}
\caption{Moments of the phase-separated density PDFs for the jet-on \high and \low runs. For each snapshot and phase we report the mean $\mu_{\log n}$, standard deviation $\sigma_{\log n}$, and skewness $\mathcal{S}_{\log n}$ in each radial shell. Dashes mark cases where a given phase is absent in that shell.}\label{tab:pdf_moments_all}
\end{table*}

\end{document}